\documentclass[letter]{aa} 
\pdfoutput=1
\usepackage{graphicx}
\usepackage{txfonts}
\usepackage[]{natbib}
\usepackage{multirow}
\usepackage{amsmath}
\usepackage[colorlinks=true,linkcolor=blue,citecolor=blue]{hyperref}
\def\deg{$^{\circ}$}
\def\degb{^{\circ}}

\begin{document}

\title{Discovery of concentric broken rings at sub-arcsec separations in the HD\,141569A gas-rich, debris disk with VLT/SPHERE\thanks{Based on data collected at the European Southern Observatory, Chile, ESO programs 095.C-0381 and 095.C-0298.}}

\author{
 	C. Perrot\inst{\ref{lesia}}, 
 	A. Boccaletti\inst{\ref{lesia}}, 					
 	E. Pantin\inst{\ref{cea}}, 							
 	J-C. Augereau\inst{\ref{ipag},\ref{ipag2}},		 	 	
 	A-M. Lagrange\inst{\ref{ipag},\ref{ipag2}}, 			
 	R. Galicher\inst{\ref{lesia}}, 						
 	A-L. Maire\inst{\ref{mpia}}, 						
 	J. Mazoyer\inst{\ref{stsci}}, 						
 	J. Milli\inst{\ref{eso}}, 							
 	G. Rousset\inst{\ref{lesia}},						
 	R. Gratton\inst{\ref{inaf}}, 						
 	M. Bonnefoy\inst{\ref{ipag},\ref{ipag2}},		  	
 	W. Brandner\inst{\ref{mpia}},  						
 	E. Buenzli\inst{\ref{zurich}},  						
 	M. Langlois\inst{\ref{cral}},  						
 	J. Lannier\inst{\ref{ipag},\ref{ipag2}},  			
 	D. Mesa\inst{\ref{inaf}},  							 
 	S. Peretti\inst{\ref{geneve}},  						
 	G. Salter\inst{\ref{lam}},  							
 	E. Sissa\inst{\ref{inaf}}, 							
	G. Chauvin\inst{\ref{ipag},\ref{ipag2}}, 			 
	S. Desidera\inst{\ref{inaf}}, 						
	M. Feldt\inst{\ref{mpia}},							
 	A. Vigan\inst{\ref{lam}},							
 	E. Di Folco\inst{\ref{bordeaux},\ref{bordeaux2}}, 	
 	A. Dutrey\inst{\ref{bordeaux},\ref{bordeaux2}}, 		
 	J. P\'ericaud\inst{\ref{bordeaux},\ref{bordeaux2}}, 	
 	P. Baudoz\inst{\ref{lesia}},  						
 	M. Benisty\inst{\ref{ipag},\ref{ipag2}},				
	J. De Boer\inst{\ref{leiden}},						
	A. Garufi\inst{\ref{zurich}}, 						
	J. H. Girard\inst{\ref{eso}},					
	F. Menard\inst{\ref{ipag},\ref{ipag2}},				
	J. Olofsson\inst{\ref{mpia},\ref{valpa1},\ref{valpa2}},
 	S. P. Quanz\inst{\ref{zurich}}, 						
	D. Mouillet\inst{\ref{ipag},\ref{ipag2}},			
 	V. Christiaens\inst{\ref{chili}, \ref{liege}},		
 	S. Casassus\inst{\ref{chili}}, 						
 	J.-L. Beuzit\inst{\ref{ipag},\ref{ipag2}}, 
 	P. Blanchard\inst{\ref{lam}}, 
 	M. Carle\inst{\ref{lam}}, 
 	T. Fusco\inst{\ref{onera},\ref{lam}}, 
 	E. Giro\inst{\ref{inaf}}, 
 	N. Hubin\inst{\ref{esog}}, 
 	D. Maurel\inst{\ref{ipag},\ref{ipag2}}, 
 	O. Moeller-Nilsson\inst{\ref{mpia}}, 
 	A. Sevin\inst{\ref{lesia}}, 
 	L. Weber\inst{\ref{geneve}}
 }

\institute{
LESIA, Observatoire de Paris, PSL Research Univ., CNRS, Univ. Paris Diderot, Sorbonne Paris Cité, UPMC Paris 6, Sorbonne Univ., 5 place Jules Janssen, 92195 Meudon CEDEX, France\label{lesia}
\and Laboratoire AIM, CEA / DSM — CNRS — Univ. Paris Diderot, IRFU / SAp, F-91191 Gif sur Yvette, France\label{cea} 
\and Univ. Grenoble Alpes, IPAG, F-38000 Grenoble, France\label{ipag}
\and CNRS, IPAG, F-38000 Grenoble, France\label{ipag2}
\and Max-Planck-Institut für Astronomie, Königstuhl 17, 69117 Heidelberg, Germany\label{mpia} 
\and Space Telescope Science Institute, 3700 San Martin Dr, Baltimore MD 21218, USA\label{stsci} 
\and European Southern Observatory, Alonso de Cordova 3107, Casilla 19001 Vitacura, Santiago 19, Chile\label{eso} 
\and INAF – Osservatorio Astronomico di Padova, Vicolo dell’Osservatorio 5, 35122 Padova, Italy\label{inaf}
\and Institute for Astronomy, ETH Zurich, Wolfgang-Pauli-Strasse 27, CH-8093 Zurich, Switzerland\label{zurich}
\and CNRS/CRAL/Observatoire de Lyon/Univ. de Lyon 1/Ecole Normale Supérieure de Lyon, Lyon, France\label{cral}
\and Geneva Observatory, Univ. of Geneva, Ch. des Maillettes 51, 1290, Versoix, Switzerland\label{geneve}
\and Aix Marseille Univ., CNRS, LAM - Laboratoire d’Astrophysique de Marseille, UMR 7326, 13388, Marseille, France\label{lam}
\and Univ. Bordeaux, Laboratoire d'Astrophysique de Bordeaux, UMR 5804, F-33270, Floirac, France\label{bordeaux}
\and CNRS, LAB, UMR 5804, F-33270 Floirac, France\label{bordeaux2}
\and Leiden Observatory, Leiden Univ., PO Box 9513, 2300 RA Leiden, The Netherlands\label{leiden}
\and Instituto de Física y Astronomía, Facultad de Ciencias, Univ. de Valparaíso, Av. Gran Bretaña 1111, Playa Ancha, Valparaíso, Chile\label{valpa1}
\and ICM nucleus on protoplanetary disks, Univ. de Valparaíso, Av. Gran Bretaña 1111, Valparaíso, Chile\label{valpa2}
\and Departamento de Astronomía, Universidad de Chile, Casilla 36-D, Santiago, Chile\label{chili} 
\and Département d’Astrophysique, Géophysique et Océanographie, Univ. de Liège, Allée du Six Août 17, B-4000 Liège, Belgique\label{liege}
\and ONERA, The French Aerospace Lab BP72, 29 avenue de la Division Leclerc, 92322 Châtillon Cedex, France\label{onera}
\and European Southern Observatory, Karl Schwarzschild St, 2, 85748 Garching, Germany\label{esog}
}

 \offprints{C. Perrot, \email{clement.perrot@obspm.fr} }


\authorrunning{C. Perrot et al.}
\titlerunning{Discovery of concentric broken rings inside 1'' in the transition disk HD 141569A with VLT/SPHERE.}

\abstract
	{Transition disks correspond to a short stage between the young protoplanetary phase and older debris phase. Along this evolutionary sequence, the gas component disappears leaving room for a dust-dominated environment where already-formed planets signpost their gravitational perturbations.}
	{We endeavor to study the very inner region of the well-known and complex debris, but still gas-rich disk, around HD\,141569A using the exquisite high-contrast capability of SPHERE at the VLT. Recent near-infrared (IR) images suggest a relatively depleted cavity within $\sim$200\,au, while former mid-IR data indicate the presence of dust at separations shorter than $\sim$100\,au.}
	{We obtained multi-wavelength images in the near-IR in J, H2, H3 and Ks bands with the IRDIS camera and a 0.95-1.35\,$\muup$m spectral data cube with the IFS. Data were acquired in pupil-tracking mode, thus allowing for angular differential imaging.}
	{We discovered several new structures inside $1''$, of which the most prominent is a bright ring with sharp edges (semi-major axis: $0.4''$) featuring a strong north-south brightness asymmetry. Other faint structures are also detected from $0.4''$ to $1''$ in the form of concentric ringlets and at least one spiral arm. Finally, the VISIR data at 8.6\,$\muup$m suggests the presence of an additional dust population closer in. Besides, we do not detect companions more massive than 1-3 mass of Jupiter.}
	{The performance of SPHERE allows us to resolve the extended dust component, which was previously detected at thermal and visible wavelengths, into very complex patterns with strong asymmetries ; the nature of these asymmetries remains to be understood. Scenarios involving shepherding by planets or dust-gas interactions will have to be tested against these observations.}	
 
\maketitle

\section{Introduction}

Observing the short phase of transition between gas-rich protoplanetary disks and dust-dominated debris disks is crucial to constrain the time when planets start to form as well as the environmental conditions.
HD\,141569A is a young \citep[5 Myr;][]{Merin2004} Herbig Ae/Be star classified as A0Ve star (V=7.12, H=6.861, K=6.821), which is located at $116_{-8}^{+9}$\,pc \citep{vanLeeuwen2007}\footnote{All distances and radii in this paper are given assuming this revised star distance (100\,mas = 11.6\,au).}. 
An optically thin disk was resolved in scattered light with Hubble Space Telescope (HST), in the near-IR, as a two-ring system located at about $\sim$250\,au and $\sim$410\,au from the star \citep{Augereau1999, Weinberger1999}.
Using HST in the visible, both \citet{Mouillet2001} and \citet{Clampin2003} observed a more complex environment  made of multiple rings and outer spirals, whose presence could be the result of an interaction with two visual stellar companions to HD\,141569A \citep{Augereau2004, Ardila2005}, with outer planets \citep{Wyatt2005}, or both \citep{Reche2009}.
From the ground, high contrast images in the near-IR were obtained with Near-Infrared Coronagraphic Imager \citep[NICI;][]{Biller2015, Mazoyer2016}, which started to probe the very inner part of the disk inside the formerly known, innermost ring at $\sim$250\,au.

While many of the structures observed in the dust distribution are representative of debris disks, HD\,141569A also contains a large amount of gas \citep{Brittain2003, Dent2005}. \citet{Thi2014} show that the gas component detected with the Herschel's PACS instrument in OI and CII cooling lines remains a major component in an hybrid disk such as HD\,141569A. 
At longer wavelengths, the CO gas component has been resolved with the IRAM's Plateau de Bure interferometer \citep{Pericaud2015} and implies a large amount of cold gas, extending out to a radius of 250\,au.

The inner region of the disk, inside $\sim$100\,au, is poorly known and is obviously of great importance when it comes to studying planetary formation and disk evolution.
Several observational facts indicate the presence of an inner dust population. First of all, the spectral energy distribution shows an IR excess at 10\,$\muup$m \citep{Thi2014} and a significant fraction of the total IR disk luminosity arises from regions closer than 100\,au \citep{Augereau1999}. 
Secondly, a resolved polycyclic aromatic hydrocarbons emission feature localized within $\sim$50\,au and about six times brighter than the expected stellar flux, has been resolved with the VLT Imager and Spectrometer for mid-InfraRed (VISIR) at 8.6\,$\muup$m \citep{Thi2014}. 
In addition, a CO gas emission line is also resolved within $\sim$50\,au by \citet{Goto2006} with an inner clearing cavity inside 10\,au. The gas kinematics indicates that the central part of the disk rotates clockwise and the southeast side is in the front, in agreement with what is inferred from the outer part \citep{Dutrey2004}.
Attempts in scattered light with differential polarimetry was unsuccessful \citep[][for instance]{Garufi2014}.
But very recently, \citet{Konishi2016} finally detected an extended disk component in the range 46 - 116\,au, corroborated by the north-south emission reported by \citet{Currie2016} in the L' band ($3.778\,\muup m$) and located at 30 - 40\,au.

This paper presents the first Spectro-Polarimetric High-contrast Exoplanet REsearch (SPHERE) observations of the HD\,141569A system angularly resolving the scattered light emission inside 200\,au in the form of several ringlets and spirals.  Section \ref{sec:obs} describes the observations and data reduction. 
We successively present the morphology of the newly resolved structures (section \ref{sec:morpho}), photometry of the brightest ringlet and a comparison with VISIR data (section  \ref{sec:iremission}), and detection limits of point-sources (section \ref{sec:pointsource}).

\section{Observations and data reduction}
\label{sec:obs}

The extreme adaptive optics coronagraphic instrument SPHERE \citep{Beuzit2008, Fusco2014} installed at the VLT in 2014, is dedicated to the search and characterization of young planetary systems. 
HD\,141569A was observed on May 2015, as part of Guaranteed Time Observation (GTO), using the Dual Band Imaging mode \citep[DBI;][]{Vigan2010} of the Infra-Red Dual-beam Imager and Spectrograph \citep[IRDIS;][]{Dohlen2008}, with filters H2 and H3. Simultaneously, a spectral data cube was obtained with the near-IR Integral Field Spectrograph \citep[IFS;][]{Claudi2008} in YJ mode ($0.95 - 1.35 \muup$m, in 39 channels). 
A second observation in open time (095.C-0381) was performed on July 2015 with IRDIS in Classical Imaging \citep[CI;][]{Langlois2014} in broadband filters J and Ks (Table \ref{obslog}). 
All observations were obtained with the Apodized Lyot Coronagraph \citep[mask diameter: 185mas,][]{Boccaletti2008}. Conditions were good for H2H3, YJ, and Ks bands and rather poor for J band ($\tau_0$\footnote{coherence time of the atmospheric turbulence at 0.5\,$\muup$m.} : 3.5, 1.1 and 0.9 ms, seeing : $0.76''$, $1.36''$ and $1.28''$, respectively for H2H3-YJ, Ks and J). 
IRDIS has a pixel size of $12.25 \pm 0.02$mas and a field-of-view (FoV) of $11'' \times 12.5''$. IFS pixel size is $7.46 \pm 0.02$mas for a $1.73'' \times 1.73''$ FoV. 
The field orientation of IRDIS and IFS are derived from astrometric calibrations as described in \citet{Maire2015}. True North corrections are given in Table \ref{obslog}. 

All the data were reduced with the SPHERE pipeline \citep{Pavlov2008} implemented at the SPHERE Data Center together with additional tools. This includes dark and sky subtraction, bad-pixels removal, flat-field correction, anamorphism correction \citep{Maire2015}, and wavelength calibration\footnote{for the IFS spectral channels as well as the transmissions of both IFS and IRDIS spectral channels.}. The location of the star is identified with the four symmetrical satellite spots generated from a waffle pattern on the deformable mirror \citep{Marois2006b}. Then, to remove the stellar halo and to achieve high contrast, the data were processed with two high-level processing pipelines : \texttt{SpeCal}, which was developed for the SPHERE survey (R. Galicher, private communication), and the processing pipeline from our team \citep{Boccaletti2015}, both leading to very similar results. We used a variety of Angular Differential Imaging algorithms: cADI \citep{Marois2006}, LOCI \citep{Lafreniere2007}, TLOCI \citep{Marois2014}, and KLIP \citep{Soummer2012}. IFS spectral frames were collapsed to increase the signal-to-noise ratio (S/N).

\begin{figure*}[t] 
\centering
\includegraphics[width=6cm]{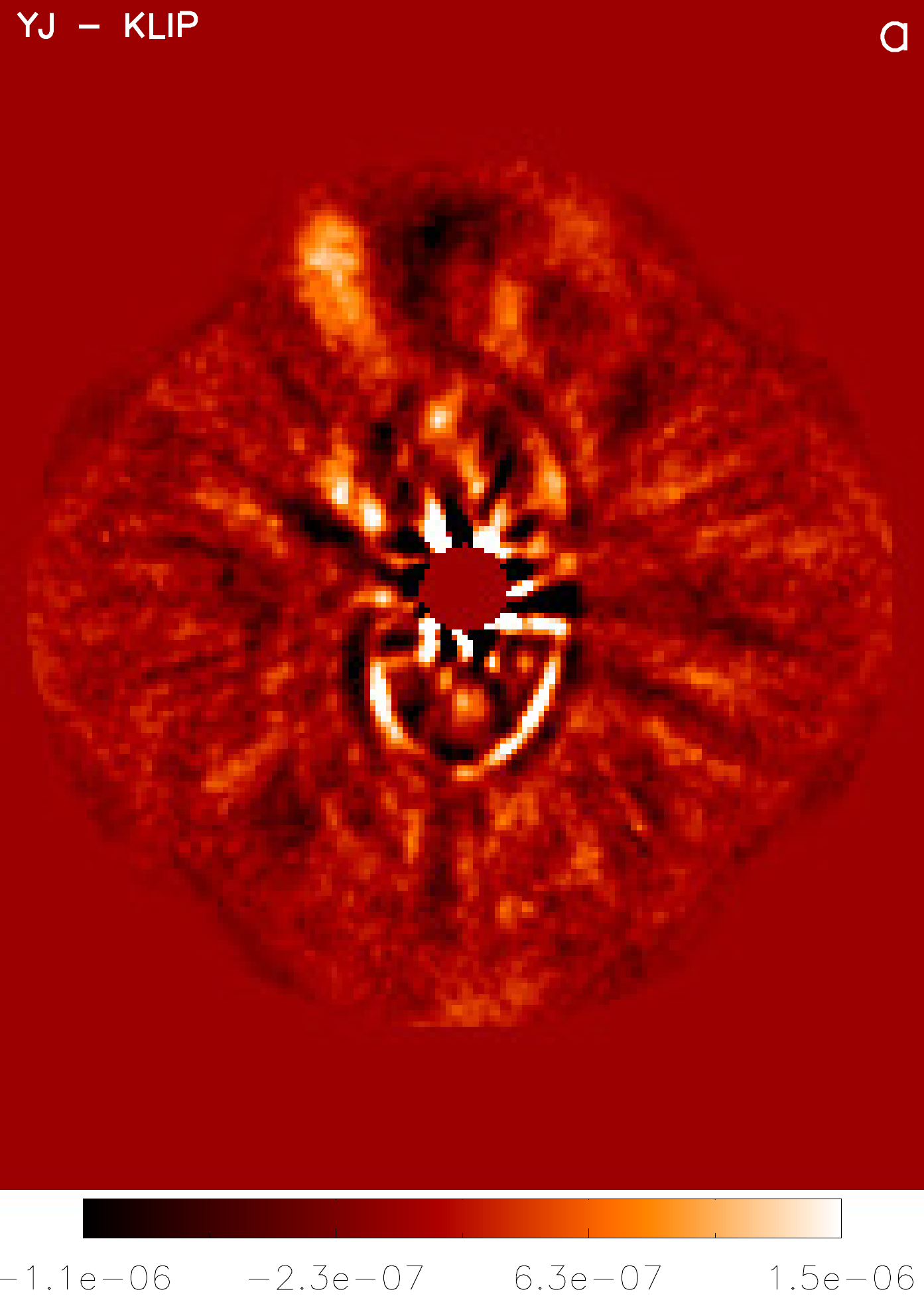}
\includegraphics[width=6cm]{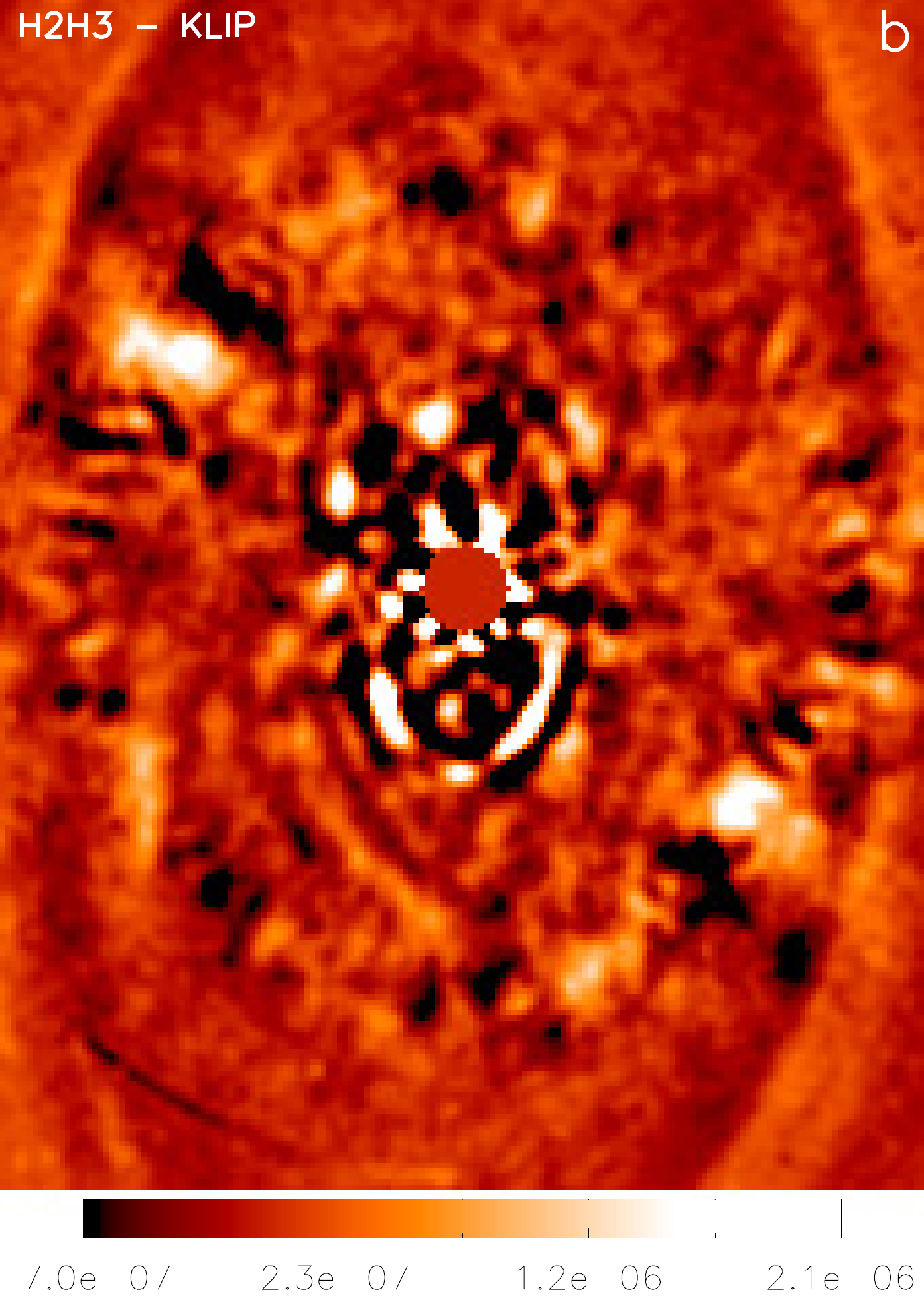}
\includegraphics[width=6cm]{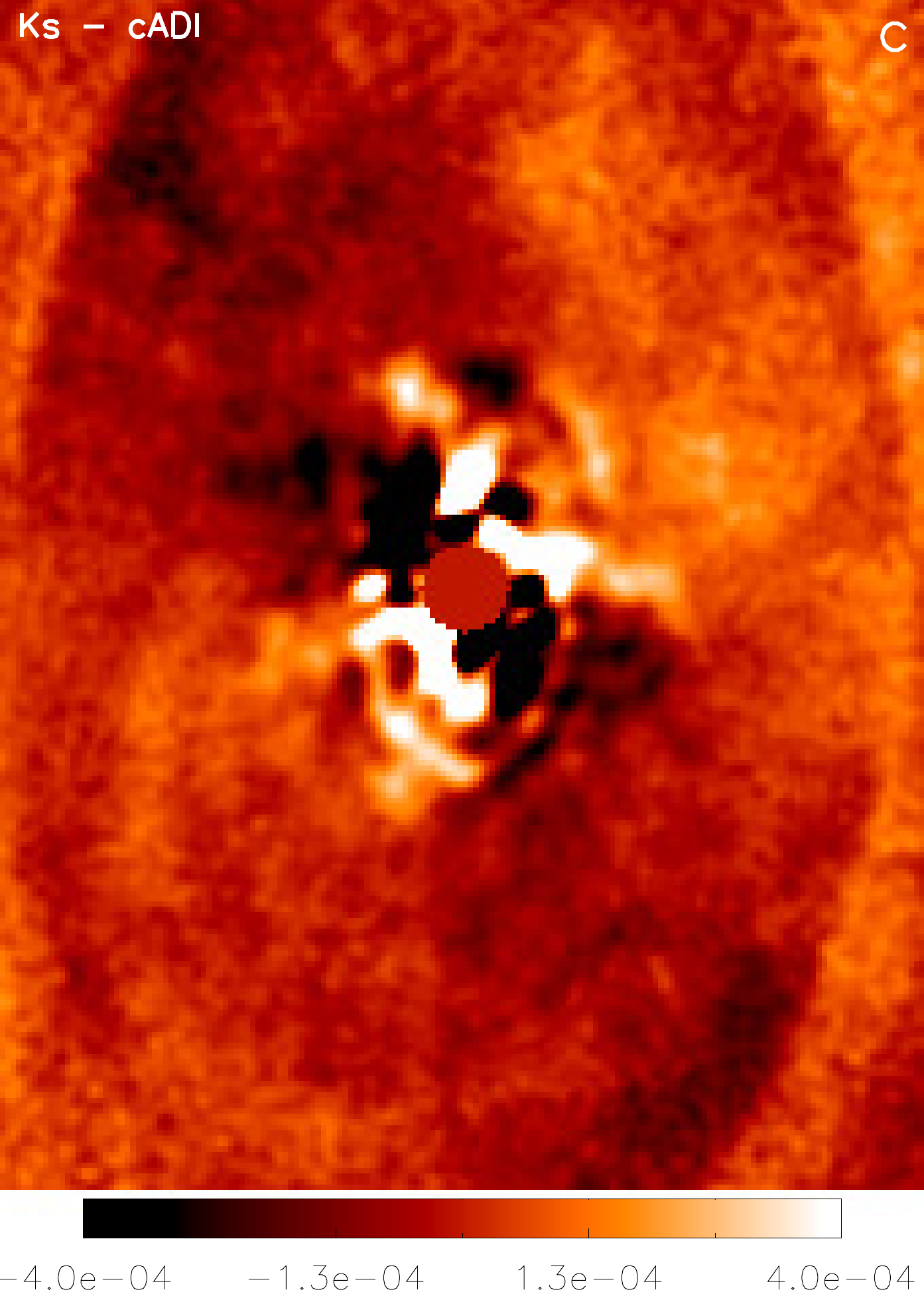}
\caption{Newly discovered structures in the inner region around HD141569A. Left : IFS - YJ (KLIP reduction). Middle : IRDIS - H2H3 (KLIP reduction). Right : IRDIS - Ks (cADI reduction). The cADI image is multipled by $r$ to improve the visibility of structures. East is left, north is top. See annotations of main structures in Fig \ref{imgannot}.}
\label{labelimg}
\end{figure*}

\section{Morphology of the inner disk}
\label{sec:morpho}

\subsection{Structures identification}

The disk is oriented at a position angle PA = 356.1$\pm$0.4\deg\ and the inclination is 56.9 $\pm$ 1.0\deg (Table \ref{fitell}); these values are in agreement with \citet{Mazoyer2016}. The best images are provided by the IRDIS instrument in H2H3 (sum of H2 and H3 images), which best compromises contrast and sensitivity (Fig. \ref{labelimg}).
All features, recently identified in \citet{Biller2015} and \citet{Mazoyer2016} are recovered at all bands (Fig. \ref{imglarge}, a) : 1) the outer ring at a semi-major axis of 3.55$''$($\sim$410\,au); 2) a very complex belt at  2.20$''$($\sim$255\,au) split in two parts, mostly on the east side; 3) an intermediate feature (either an arc or a broken spiral arm) extending to the east from north to south in between these two rings, which clearly roots at the top of the aforementioned belt; 4) finger-like features at 1.35-1.90$''$, PA$\approx$148\deg; and 5) a very dark and steep inner cavity inwards of $1.84"$ ($\sim$210\,au).

In addition, the SPHERE images provide deep insight into the central $\sim$200\,au region. We identify several new structures that are either spiral-like or ring-like.
A spiral arm S1, detected in Ks, starts as close as $\sim$0.5$''$ to the east of the star and winds to the southeast (Fig. \ref{labelimg} and \ref{labelimg} \& \ref{imglarge}, b for annotations).
A more extended region is visible as a possible counterpart to the north and wrapping to the east (S2 ?, \ref{labelimg} and \ref{labelimg} \& \ref{imglarge}, b for annotations). 
Closer in, a series of at least three structures resembling ringlets (labelled R1 to R3, Fig. \ref{labelimg} and \ref{imgannot} for annotations) and featuring strong asymmetries and clumpiness are discovered from both IRDIS (H2H3, Ks) and IFS (YJ) data.
The J band data, owing to poor observing conditions, do not achieve large enough contrasts to detect these new features. The main pattern R3 is also the brightest part of  the whole disk, especially if we take into account that closer in the ADI attenuation is larger \citep{Milli2012}. The R3 pattern is strongly asymmetrical, mostly visible in the southern part while the northern part is just barely detected ($\sim$5 times fainter than the southern part).
Surprisingly, this is orthogonal, hence, inconsistent, with the global east-west inclination of the disk, in which forwards scattering can create a brightness asymmetry with respect to the semi-major axis.
We cannot posit a pericentre glow effect \citep{Wyatt2005} because R3 appears centred onto the star, following the north-south direction. Therefore, we suspect a true depletion of dust in this ring towards the north. The mechanisms that can cause such large azimuthal variations of the dust density remain to be studied. A clump, nearly point-like is visible in the southern ansae of R3 ($PA\sim178$\deg, $r\sim0.41$'') surrounded by a drop of intensity on each side (Fig. \ref{labelimg} and \ref{imgannot} for annotation), where the eastern side is darker. 
This feature could be a consequence of ADI artifact because a similar structure appears in forwards modelling of featureless synthetic dusty disks (see section \ref{sec:iremission} and Fig. \ref{model}). The clump and drop of intensity, however, are still detected in a spectral TLOCI reduction of the IFS data, which is not affected by the ADI bias. Moreover, the clump is slightly shifted with respect to the ansae of R3.
Whether it could be associated with a real object requires more data that do not suffer from ADI biases (polarimetry for instance). Two other much fainter structures which that look like broken rings, R1 and R2, are visible at larger separations than R3 (Fig. \ref{labelimg} and \ref{imgannot} for annotations). 
All these three ringlets are recurrent patterns in the various datasets YJ, H2H3, and Ks, collected at two epochs (Fig. \ref{imgannot}). Their elliptical shape departs from the nearly circular starlight residuals, which are particularly strong at the correction radius ($\sim 0.8 ''$ in H band). Given that the disk is visible in a large range of radius and azimuthal angle, calculating a S/N map to test the reliability of R1-R3 would be impractical. 
Instead we plot the radial profiles of the KLIP-H2H3 deprojected image, azimuthaly averaged in four quadrants (Fig. \ref{coupe}). The ringlets R3 and R2 are clearly identified as bumps localized at a constant radius. The case of R1 is more ambiguous as it appears in only two quadrants (SE and NW) and at different radii. Therefore it is not ascertained whether R1 is a ringlet or a spiral.
Other fainter structures (also elliptical) may possibly exist but are not differentiable from speckles.
It is yet unclear whether these three rings correspond to a nearly concentric system, which to some extent are similar to those around HL Tau \citep{ALMA2015} and TW Hydrae \citep{Rapson2015}, or wherther they are hints of spiral arms.

The extended and nearly continuous disk component detected by HST's STIS instrument \citep{Konishi2016} is not visible in the SPHERE images partly because the ADI process filters out such broad features. Instead, we are sensitive to higher frequency variations on top of this inner disk which we resolve as ringlets or spirals.
The $L'$ detection from \citet{Currie2016} partially matches with the ansa of R3 and the structures they named $H1$ and $H2$ may correspond to R2.
However, they did not observe any north-south asymmetry visible at shorter wavelengths with SPHERE.
Moreover, the point-like source reported by \citet{Currie2016} ($PA \sim 180$\deg, $R \sim 0.28 ''$) is not detected in our images (see detection limits in section \ref{sec:pointsource}).

\subsection{Localization of the structures}
Structures were registered in a similar way as in \citet{Boccaletti2013}. 
First, we extracted the radial profiles (azimuthally sampled by steps of 1\deg) of structures that are detected in several wavelengths.
A one-dimensional (1D) Gaussian model is fitted on these profiles to provide the location of the maxima of the structures. 
These measurements for R1, R2, and R3 are reported in Fig. \ref{imgmesure}.
Assuming the ringlets are each part of an individual inclined ring, 
we used a non-linear least squares algorithm to fit these maxima with an elliptical contour considering a Gaussian weighting.
The free parameters of the elliptical contour are the semi-major and semi-minor axes, the position angle (PA) and offsets with respect to the position of the star. We performed the fitting for each band (IRDIS Ks and H2H3, IFS YJ) and each algorithm (cADI, TLOCI, and KLIP). 
Table \ref{fitell} provides averaged values and dispersions for different algorithms and wavelengths. 

We found that all ringlets R1, R2, and R3 have an inclination in the range $56-58$\deg compatible with the inclination found for the inner ring (56.9$\pm$1.0\deg) within error bars. The  PAs of R2 and R3 are slightly different than the global orientation of the disk by $\sim 1-2$\deg. 
Finally, we measured offsets of $15.4\pm 3.4$\,mas ($1.79\pm0.40$\,au), $15.4\pm4.8$\,mas ($1.79\pm0.56$\,au), and $34.9\pm5.1$\,mas ($4.05\pm0.59$\,au) towards west, respectively for R3, R2, and R1, plus an offset of $82.2 \pm 17.0$\,mas ($9.54\pm1.97$\,au) towards north for R1.  
In addition, the inner ring at 210\,au has an opposite offset direction. We measured $29.8 \pm 7.8$\,mas ($3.46\pm0.90$\,au) towards east and $32.8 \pm 6.9$\,mas ($3.80\pm0.80$\,au) towards north in agreement with \citet{Mazoyer2016}.
Here we did not considered ringlets ellipticity even though could account for the differential offset. In particular, R1 has an important offset (Table \ref{fitell}). As explained above, however, the exact nature of R1 (ringlet or spiral) is left undetermined.

Considering the linear wave density theory \citep{Rafikov2002} and the tools we previously developed \citep{Boccaletti2013}, we attempted to fit the spiral feature S1 on the deprojected disk image. We did not find a set of parameters which produces a match between the model and S1 using a simpler Archimedean spiral model either. 
A more sophisticated model might be required to account for the shape of S1 if produced by a planet \citep{Dong2015a}. Alternatively, the system could have experienced gravitational instabilities \citep{Dong2015b}, or the spiral arm could be in a different plane than the other parts of the disk, as suggested in \citet{Biller2015}.

\section{Photometry}
\label{sec:iremission}

We performed disk modelling and photometry of the ringlet R3 to derive qualitative results (a thorough modelling is postponed to future work). Following earlier work \citep{Boccaletti2012, Mazoyer2014}, we built a set of geometrical disks model using \texttt{GRaTer} \citep{Augereau1999b} with no particular assumption about the grain optical properties. To restrain the parameter space we fixed the PA of the ringlet to  356.5\deg\, (compatible with the fit previously described). The inclination $i$ and the position of the ring ($r_0$) span a narrow range of values ($i$ = 57, 58, 59\deg; $r_0$ = 47, 48, 49\,au) since R3 is already well defined in the images. The model assumes that the surface density radially decreases inwards and outwards of $r_0$ as power laws with slopes $\alpha_{in}$ and $\alpha_{out}$, respectively.  We used a large range of values for these parameters: 2, 5, 10 and 20 for $\alpha_{in}$ and -2, -5, -10 and -20 for $\alpha_{out}$.
We set the aspect ratio to $h/r=0.01$ \citep[following][]{Thi2014}, where $r$ is the separation to the star and $h$ is the height of the disk. Moreover, we assumed a front-back (east-west) symmetrical ring with isotropic scattering ($g=0$) as we are interested in the southern part of R3 alone.
We ran the least squares minimization between the grid of models and the IRDIS H2H3 KLIP image. The best model (minimum $\chi^2$) yields $i=57\degb$ and  $r_0$ = 48\,au (0.41$"$), which are both in close agreement with the ellipse fitting, and surface density slopes of $\alpha_{in}=20$ and  $\alpha_{out}=-20$. Considering a more conservative threshold accounting for the degree of freedom in the fitting procedure, we end up with a list of possible models all having surface density slopes modulus of 10 or 20.  Similar results are derived from model fitting in the Ks band.
We can conclude that the ringlet R3 is rather narrow (radially unresolved) and has very sharp edges both inwards and outwards (Fig. \ref{model}), which implies that it is probably bounded by perturber(s) or shaped by the coupling of gas and dust as described for example in \citet{Lyra2013}.

Using the best model to estimate the ADI bias, we measured the integrated intensity in the southern part of R3 to be 0.45\,mJy  and 0.35\,mJy in the H and Ks bands, respectively. These numbers stand for rough estimations, but are consistent with the variation of the flux of the star from H to Ks, as we should expect for scattering.

In addition, a simple radiative transfer toy model was developed to test the geometry probed by SPHERE observations against the VISIR images shown in \citet{Thi2014}.
This toy model is based on a set of concentric rings for which the radii are set according to the angularly resolved existing images in scattered light: 380-420\,au , 190-210\,au, and the newly found 45-48\,au ringlets (R3).  Using only this set of three concentric rings, the resulting profile at 8.6\,$\muup$m is  inconsistent with VISIR 8.6\,$\muup$m image. The thermal flux in the inner 0.2$''$ (equivalent to the VISIR resolution) is not large enough compared to what is observed. In the mid-IR an additional component is thus required closer to the star. Given VISIR resolution, any additional ring-shaped component with a mean radius smaller than $\sim$20\,au would be compatible with VISIR data. 

\section{Detection limits}
\label{sec:pointsource}

\begin{figure} 
\resizebox{\hsize}{!}{\includegraphics{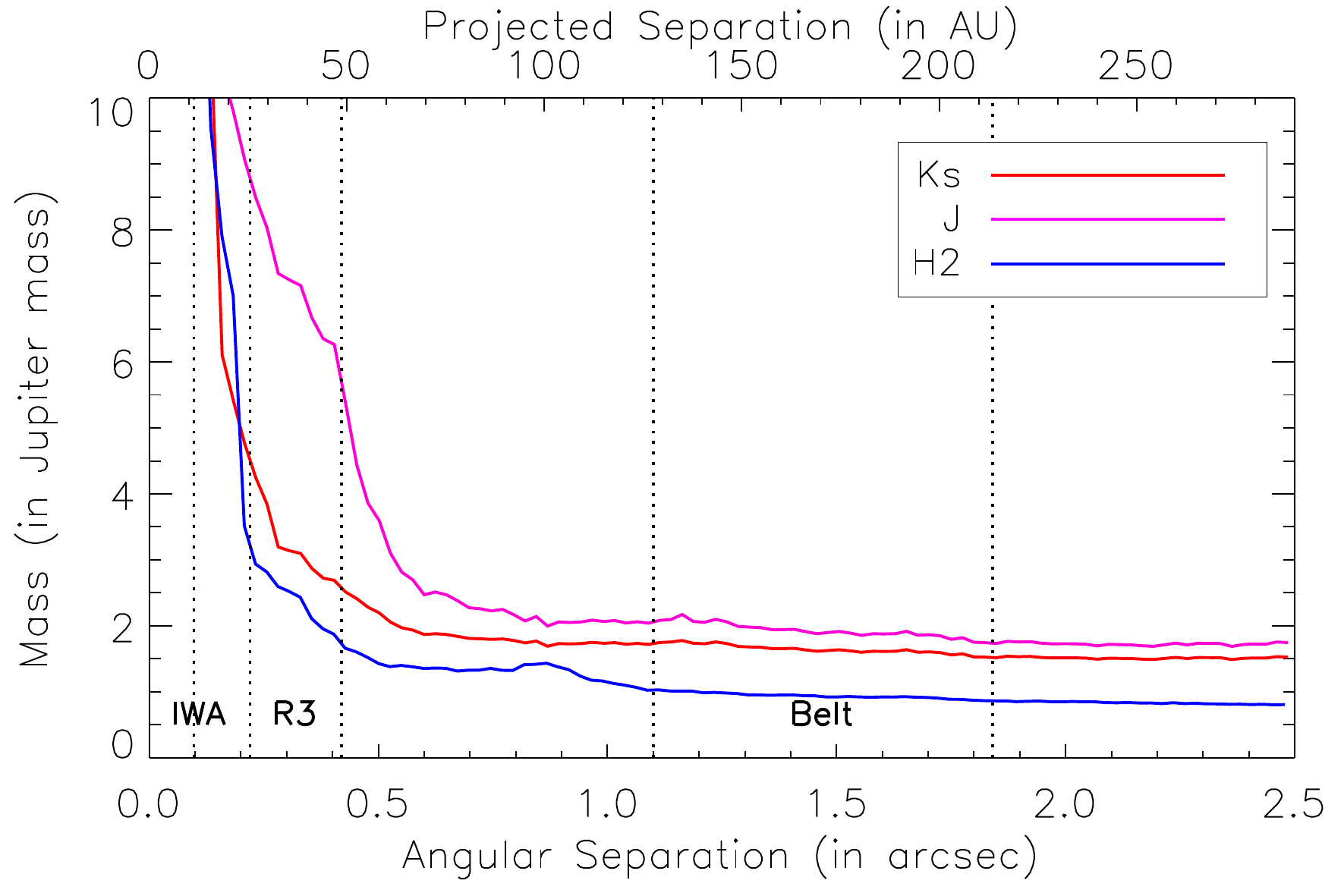}}
\caption{Detection limit in Jupiter mass for J, H2 and Ks band, assuming the BHAC-2015+COND-2003 model. IWA is the inner working angle of the coronagraph. R3 and Belt represent the separation where the structures are located.}
\label{cstcurve}
\end{figure}

While the many structures of the transition disk HD\,141569A suggest the presence of planets, we do not detect any reliable point sources, apart from speckles, 
which are present inside 1'' (for instance at $r\sim0.38''$, $PA\sim11$\deg and $r\sim0.35''$, $PA\sim51$\deg). Hence, we measured the contrast, at $5\sigma$, in J, H2H3, Ks and YJ (Fig. \ref{cstcurve1}) bands for data processed with TLOCI (optimized for point sources). The ADI throughput is accounted for by a customized pipeline, \texttt{SpeCal}. 
From these detection limits, we converted in Jupiter mass (Fig. \ref{cstcurve}), assuming the latest evolutionary model \citep[BHAC-2015 + COND-2003,][]{baraffe2003}. The IFS YJ and IRDIS H2H3 contrasts are far superior to the other bands as a result of data quality. 
For an arbitrary separation of 0.5$''$ (roughly where the detection limit in mass starts to flatten), the H2 (respectively Ks) image would have allowed the detection of 1 - 2\,$\textrm{M}_\textrm{J}$ (respectively, 2 - 3\,$\textrm{M}_\textrm{J}$).
The limit in J band is worse inside $0.6"$ but then similar to the limit in Ks outwards.
We found no planets more massive than 1 - 3\,$\textrm{M}_\textrm{J}$, between 0.3$''$ (ringlet R3) and 1.84$''$ (the edge of the inner belt). Inside the radius of R3 the detection performance degrades rapidly to about 10\,$\textrm{M}_\textrm{J}$ near the inner working angle (IWA) of the coronagraph (93\,mas).
At the PA and separation of the point-like source reported by \citet{Currie2016} (estimated to 5 - 6\,$\textrm{M}_\textrm{J}$), and not observed in our image, our data yield a detection limit of 3.5\,$\textrm{M}_\textrm{J}$ in Ks and 2.5\,$\textrm{M}_\textrm{J}$ in H2, hence, this limit is not compatible with the mass derived from L'. 

\section{Conclusion}
\label{sec:conclu}
Exploring the inner 1$''$ region with SPHERE of the transition disk HD\,141569A has revealed a series of concentric ringlets at physical separation of 47\,au, 64\,au and 93\,au, partially associated with the emission formerly detected by the VISIR instrument \citep{Thi2014}. An additional dust component closer than 0.2$''$ may be required to account for this emission.
However, these new structures match perfectly the extended disk component found by STIS \citep{Konishi2016} and the north-south extension by \citet{Currie2016}. 
We have shown that the inclination and the PA of each of these ringlets are in perfect agreement with those of the outer belts. The brightest ringlet is so asymmetrical that it appears as a half-ring. This brightness asymmetry is not consistent with forwards scattering and may then be the result of a true azimuthal variation of the dust density; the reason for this has yet to be understood. We also noted the presence of a clump in the south of the brighter ringlet, which could be an artifact from the reduction process or a local variation of the dust density. In addition, it is difficult to trace each of these broken rings exactly in all directions hence, there is a possible confusion with spiral patterns.
A large spiral pattern is observed in Ks, developing southwest with possibly a northern counterpart.
These observed structures could be the consequence of perturbations by planets, which confine dust grains and create gaps. 
However, we found no planets more massive than 1-3 $\textrm{M}_\textrm{J}$ between the edge of the cavity and the ringlet R3. 
These kinds of structures could also be created by the coupling of gas and dust triggering instabilities in the form of narrow eccentric rings when the gas-dust ratio is close to the unity \citep{Lyra2013}.

\begin{acknowledgements} 
We acknowledge financial support from the Programme National de Planétologie (PNP) and the Programme National de Physique Stellaire (PNPS) of CNRS-INSU. This work has also been supported by a grant from the French Labex OSUG@2020 (Investissements d’avenir – ANR10 LABX56).
The  project  is  supported  by  CNRS,  by  the  Agence  Nationale  de  la  Recherche  (ANR-14-CE33-0018). 
This work is partly based on data products produced at the SPHERE Data Centre hosted at OSUG/IPAG, Grenoble. We thank P. Delorme and E. Lagadec (SPHERE Data Centre) for their efficient help during the data reduction process.
SPHERE is an instrument designed and built by a consortium consisting of IPAG (Grenoble, France), MPIA (Heidelberg, Germany), LAM (Marseille, France), LESIA (Paris, France), Laboratoire Lagrange (Nice, France), INAF-Osservatorio di Padova  (Italy), Observatoire de Genève (Switzerland), ETH Zurich (Switzerland), NOVA (Netherlands), ONERA (France) and ASTRON (Netherlands) in collaboration with ESO. 
SPHERE was funded by ESO, with additional contributions from CNRS (France), MPIA (Germany), INAF (Italy), FINES (Switzerland) and NOVA (Netherlands).
SPHERE also received funding from the European Commission Sixth and Seventh Framework Programmes as part of the Optical Infrared Coordination Network for Astronomy (OPTICON) under grant number RII3-Ct-2004-001566 for FP6 (2004-2008), grant number 226604 for FP7 (2009-2012) and grant number 312430 for FP7 (2013-2016).
V.C. acknowledges support from the European Research Council under the European Union's Seventh Framework Programme (ERC grant agreement No. 337569) and from the French Community of Belgium through an ARC grant for Concerted Research Action.
J. O. acknowledges support from the Millennium Nucleus RC130007 (Chilean Ministry of Economy).

\end{acknowledgements}

\bibliography{hd141569_sph_biblio}

\clearpage
\appendix

\section{Observing log and fit tables}

\begin{table*}[h!]
\centering
\footnotesize
\begin{tabular}{ l l l l l l l l l l }
\hline
Programme  & Instrument & Filter & $\lambda_c$ & Date       & FoV Rotation & $T_{exp}$ & DIT & $N_{exp}$ & TN \\ 
		   & 			& 		& ($\muup m)$ & UT         & (\deg) 		& (s) 		& (s) &  		 & (\deg) \\ \hline
095.C-0298 & IRDIS 		& H2 	& 1.593		  & 2015-05-16 & 42.07 		& 4096 		& 64  & 64 		 & $-1.8\pm0.1$\\
095.C-0298 & IRDIS 		& H3 	& 1.667 		  & 2015-05-16 & 42.07 		& 4096 		& 64  & 64 		 & $-1.8\pm0.1$\\
095.C-0298 & IFS   		& YJ 	& 0.95 - 1.35 & 2015-05-16 & 42.07 		& 4096 		& 64  & 64 		 & $-1.8\pm0.1$\\
095.C-0381 & IRDIS 		& J 		& 1.245 		  & 2015-07-22 & 34.89 		& 3200 		& 8   & 400 		 & $-1.67\pm0.03$\\
095.C-0381 & IRDIS 		& Ks 	& 2.182		  & 2015-07-28 & 35.54 		& 3200 		& 16  & 200 		 & $-1.67\pm0.03$\\
\hline
\end{tabular}
\normalsize
\caption{Observing log: Programme name, Instrument, acquisition mode and filter, central wavelength, date, variation of parallactic angle, total exposure time, individual exposure time, number of frames, true North calibration.}
\label{obslog}
\end{table*}

\begin{table*}[h!]
\centering
\footnotesize
\begin{tabular}{ l l l l l l l }
\hline
Structure & Semi-major axis & Semi-minor axis & PA & Inclination & West offset & North offset\\
 & (mas) & (mas) & (\deg) & (\deg) & (mas) & (mas) \\ \hline
Inner ring 	& $1774.8 \pm 11.2$ & $970.1 \pm 18.6$ & $356.1 \pm 0.4$ & $56.9 \pm 1.0$ & $-29.8 \pm 7.8$ & $32.8 \pm 6.9$  \\
R1   		& $805.0 \pm 14.8$  & $431.2 \pm 7.9$  & $356.0 \pm 2.0$ & $57.6 \pm 1.3$ & $34.9 \pm 5.1$ & $82.2 \pm 17.0$\\
R2  			& $549.1 \pm 12.1$  & $307.0 \pm 11.0$ & $354.5 \pm 1.0$ & $56.0 \pm 2.2$ & $15.4 \pm 4.8$ & $5.8 \pm 4.8$\\
R3   		& $406.2 \pm 7.2$   & $215.8 \pm 3.8$  & $353.7 \pm 1.1$ & $57.9 \pm 1.3$ & $15.4 \pm 3.4$ & $1.2 \pm 9.4$\\
\hline
\end{tabular}
\normalsize
\caption{Parameters for the ringlets assuming offset ellipses.} 
\label{fitell}
\end{table*}

\section{Multi-views and annotations of HD\,141569A disk}

\begin{figure*}[h!]
\centering
\includegraphics[width=4.5cm]{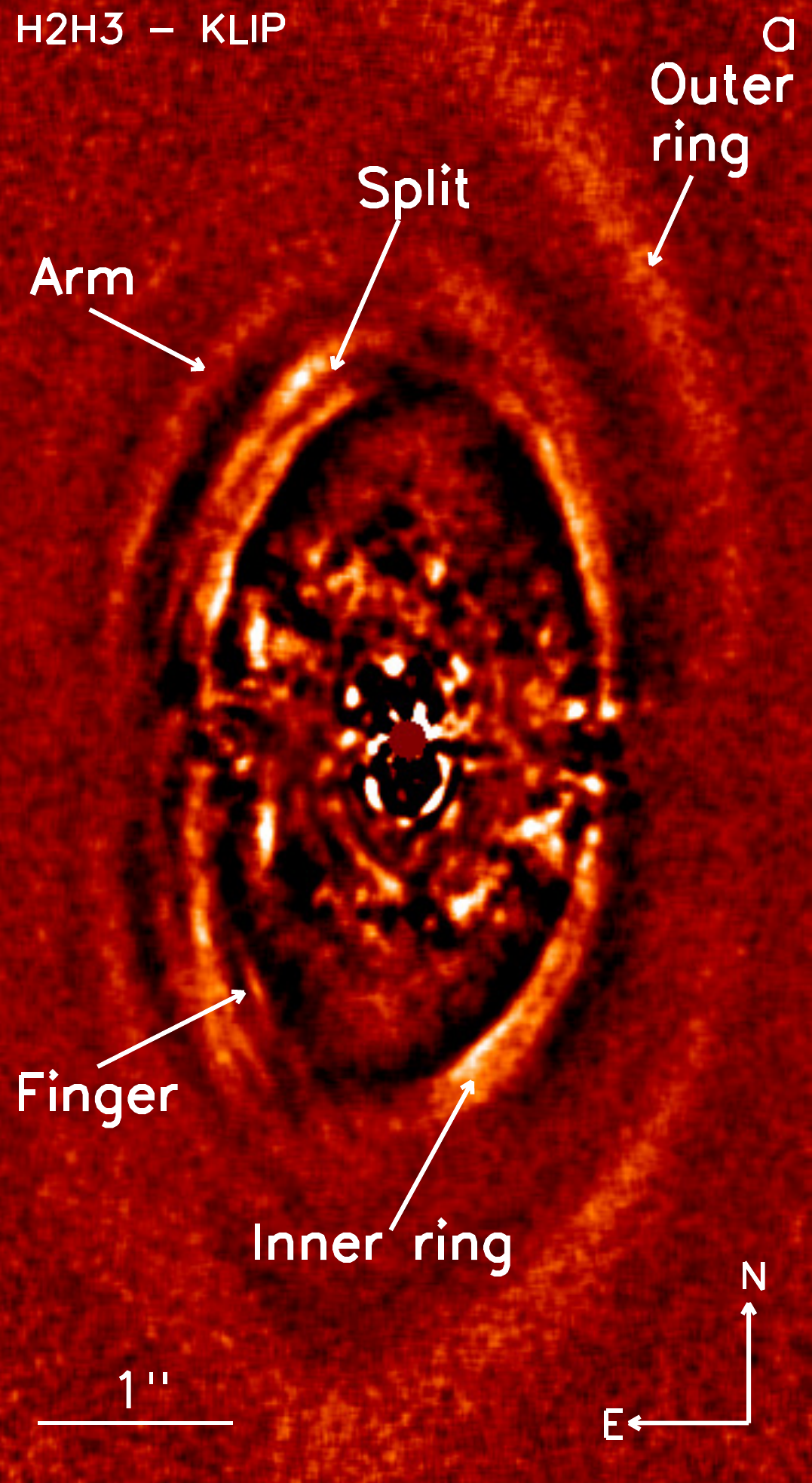}
\includegraphics[width=4.5cm]{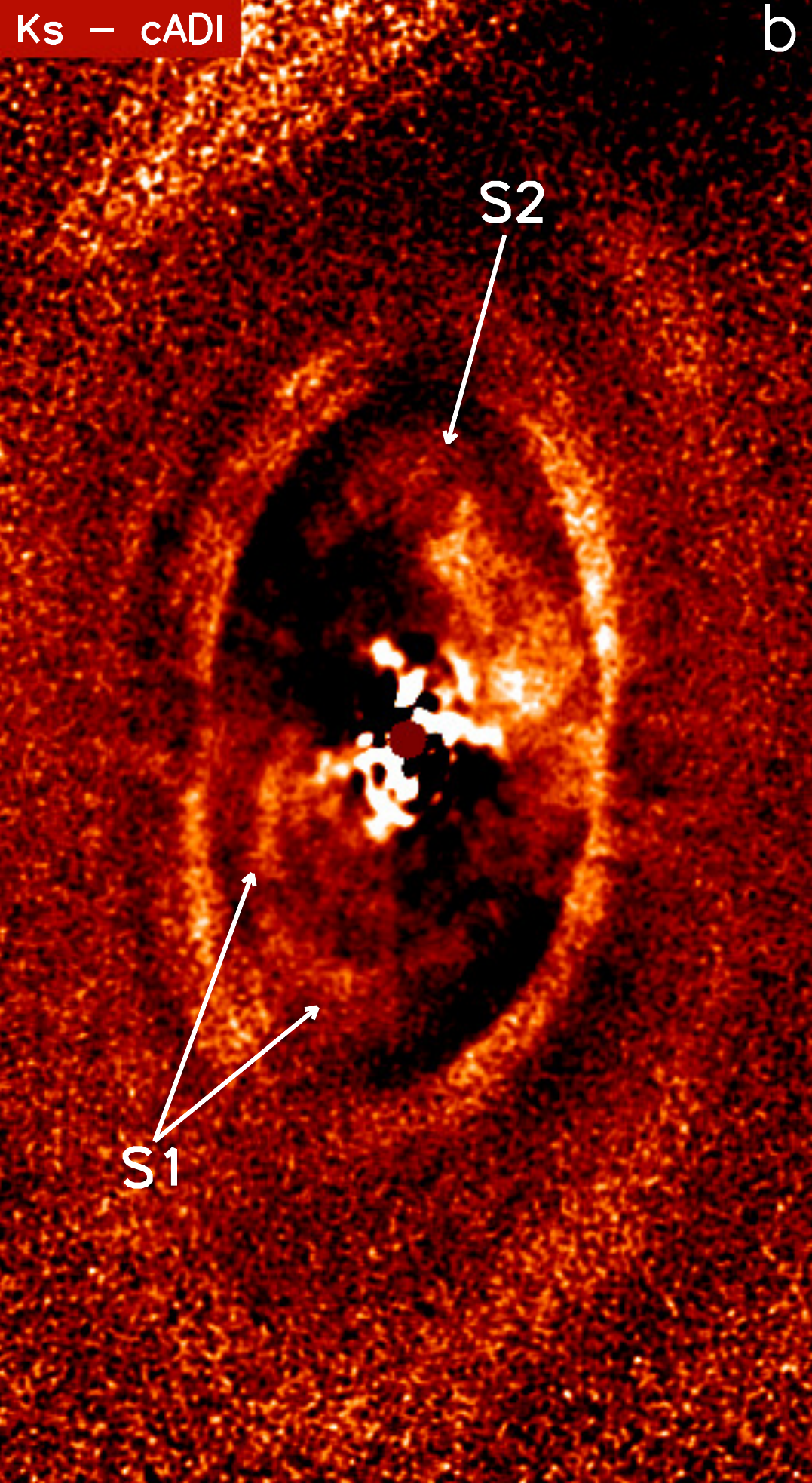}
\includegraphics[width=4.5cm]{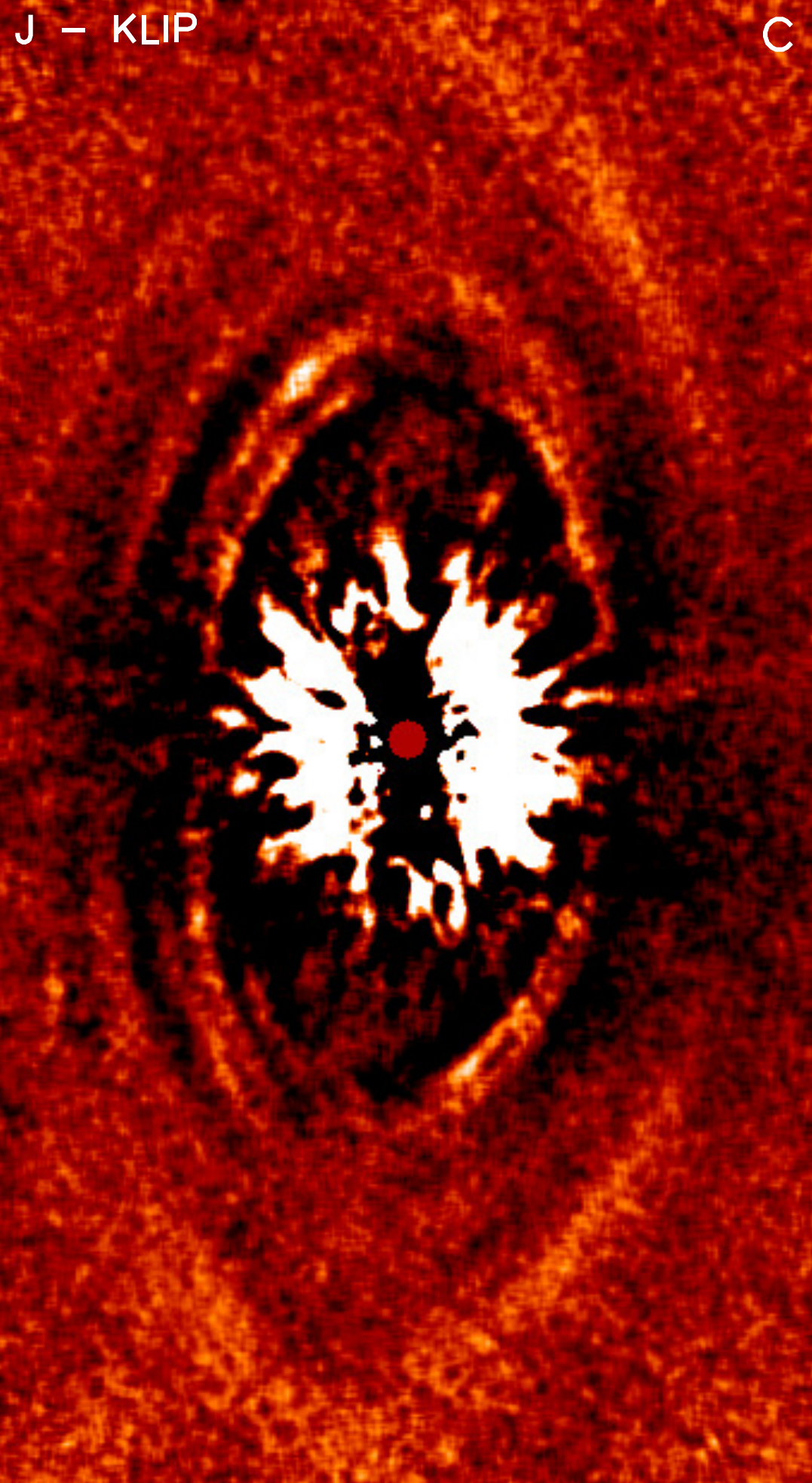}
\includegraphics[width=4.5cm]{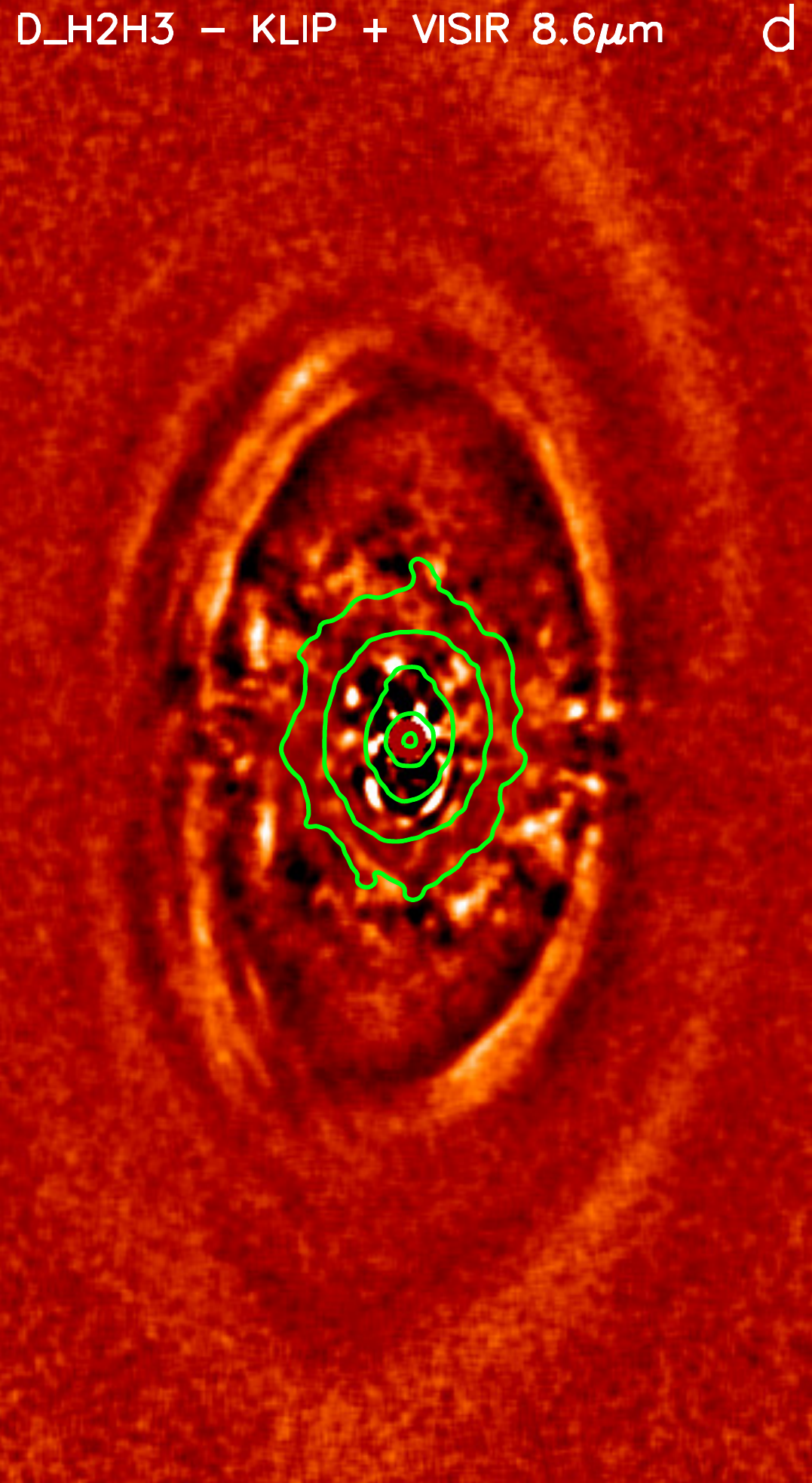}
\caption{Reduced images of the disk around HD\,141569A. a) wide field of view obtained in H2H3 \citep[KLIP reduction, annotation from][]{Mazoyer2016}. b) cADI reduction of Ks image with the annotation of the spirals S1 and S2. c) KLIP reduction of J image. d) KLIP reduction of H2H3 image with the VISIR contour at 8.6\,$\muup$m. All images are arbitrarily multiplied by the distance to the star in pixel for cosmetics reason. The spatial scale is the same for the four images. East is left, north is up.}
\label{imglarge}
\end{figure*}

\begin{figure*}[h!]
\centering
\includegraphics[width=5.5cm]{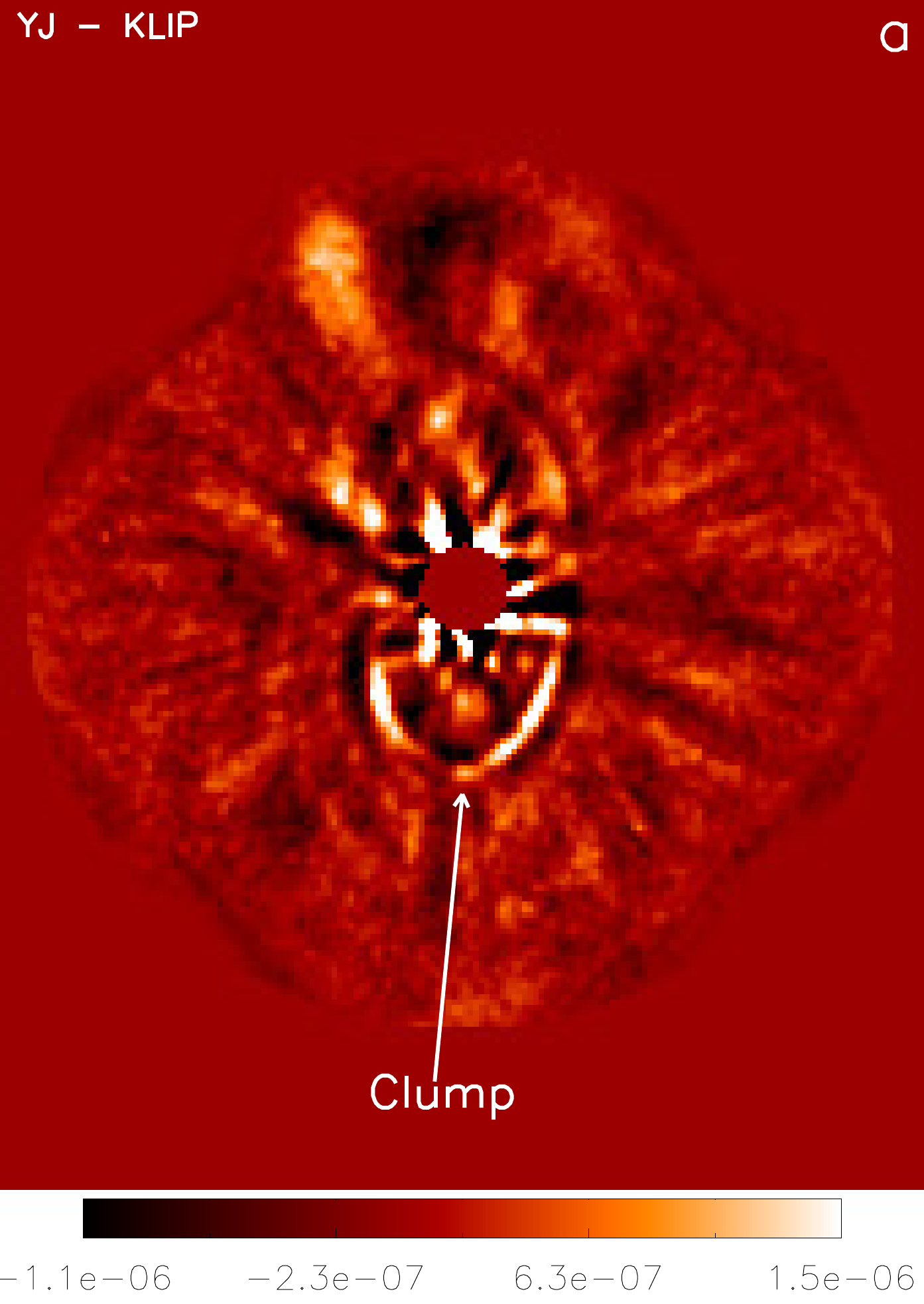}
\includegraphics[width=5.5cm]{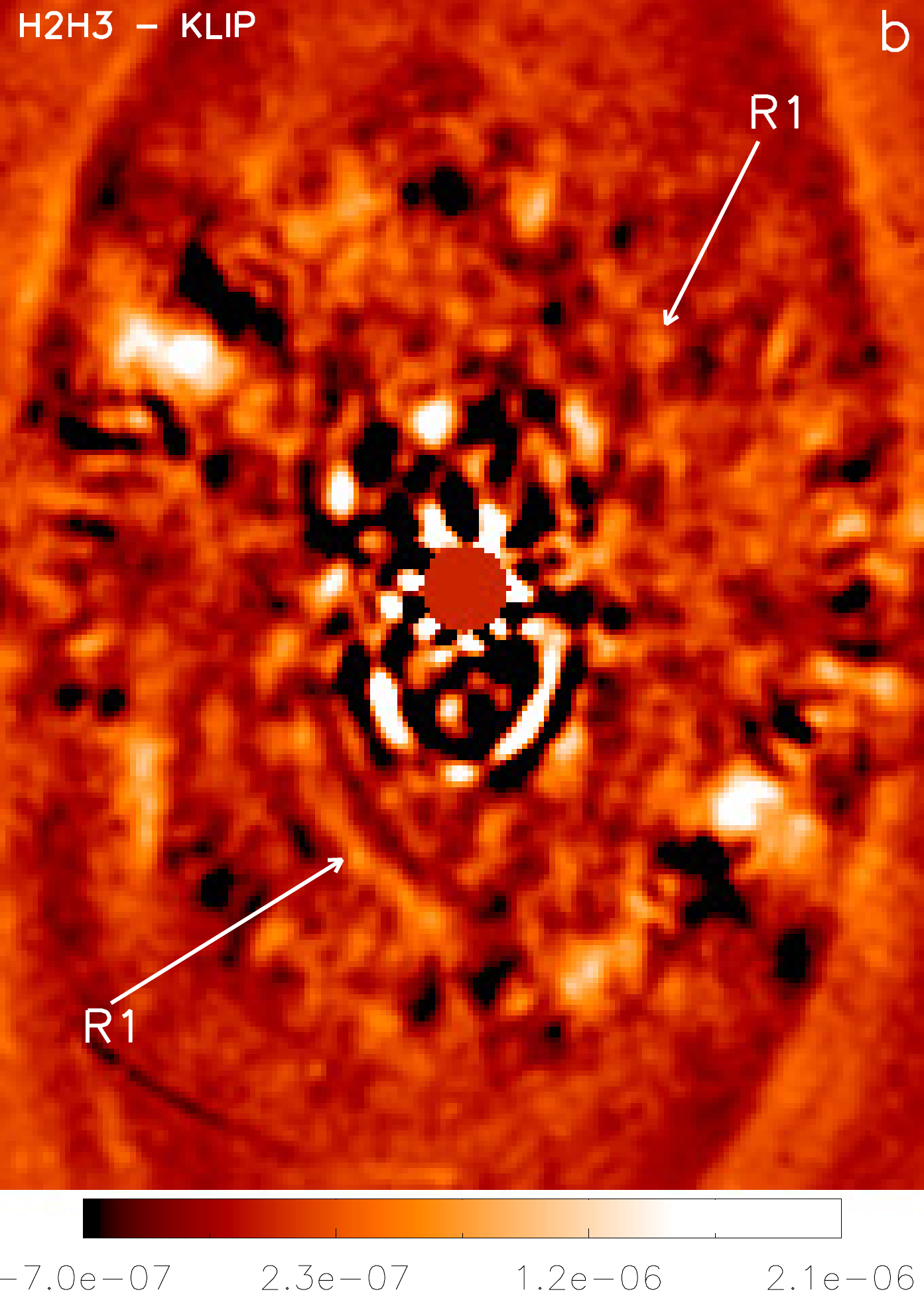}
\includegraphics[width=5.5cm]{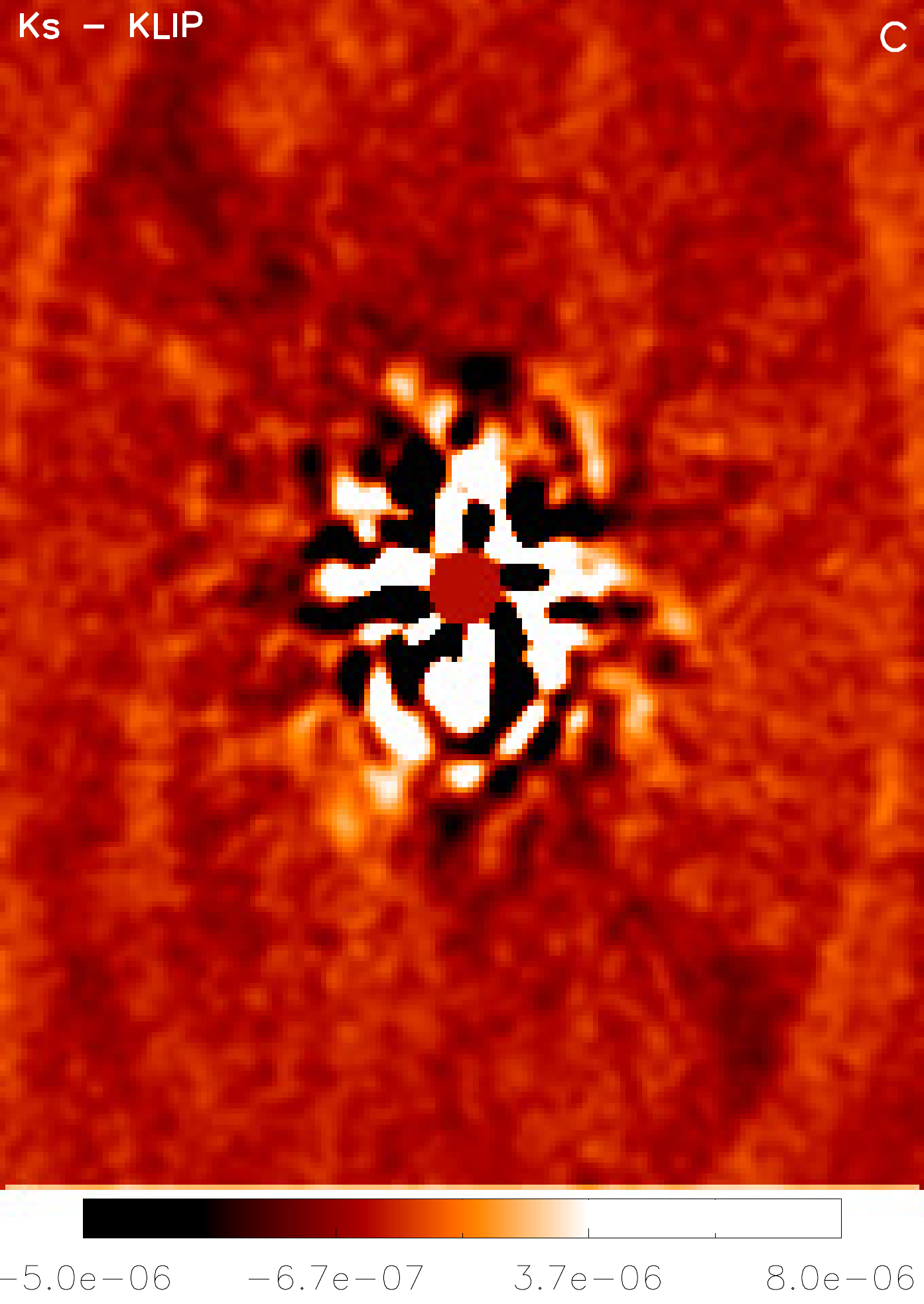}\\
\includegraphics[width=5.5cm]{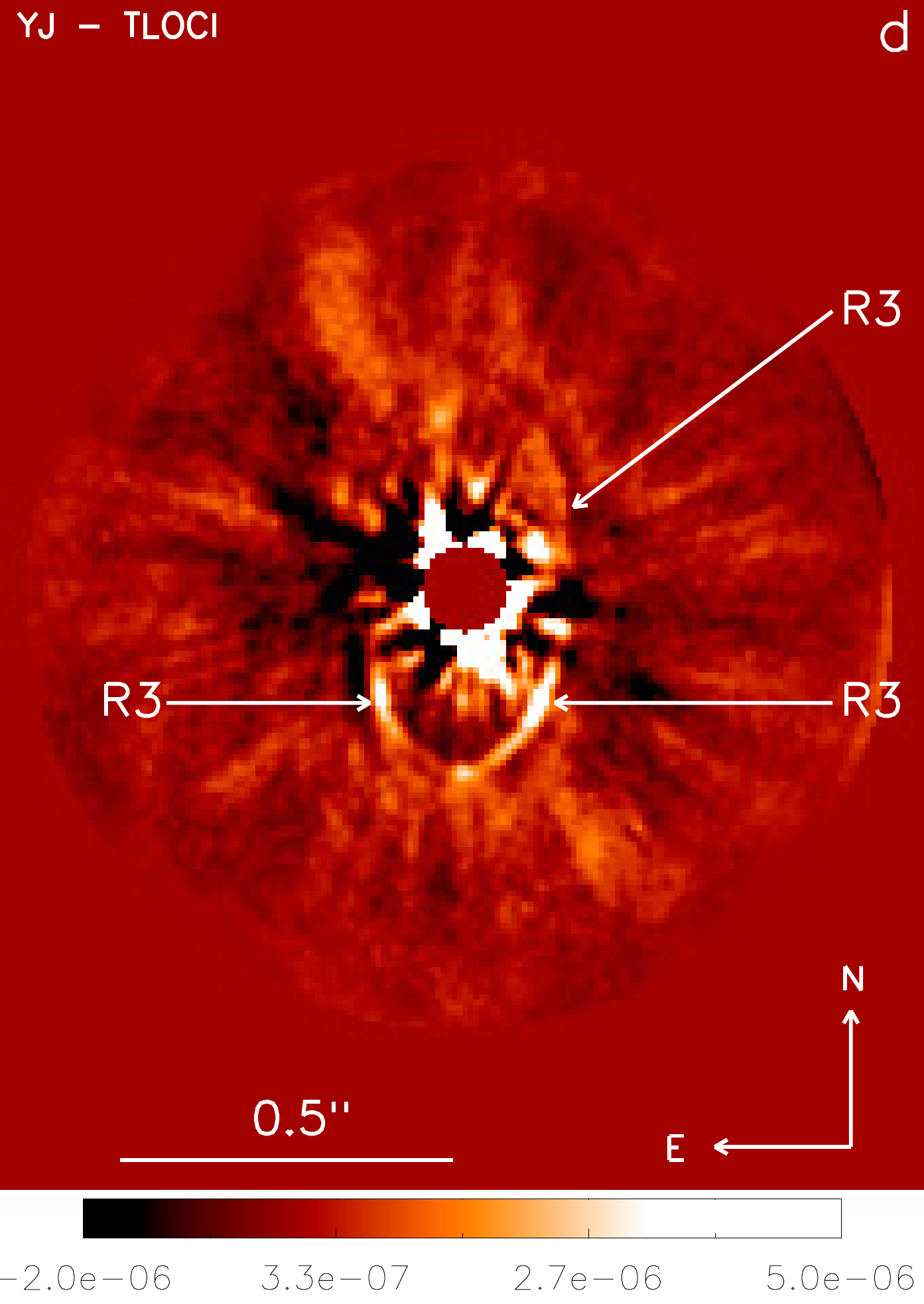}
\includegraphics[width=5.5cm]{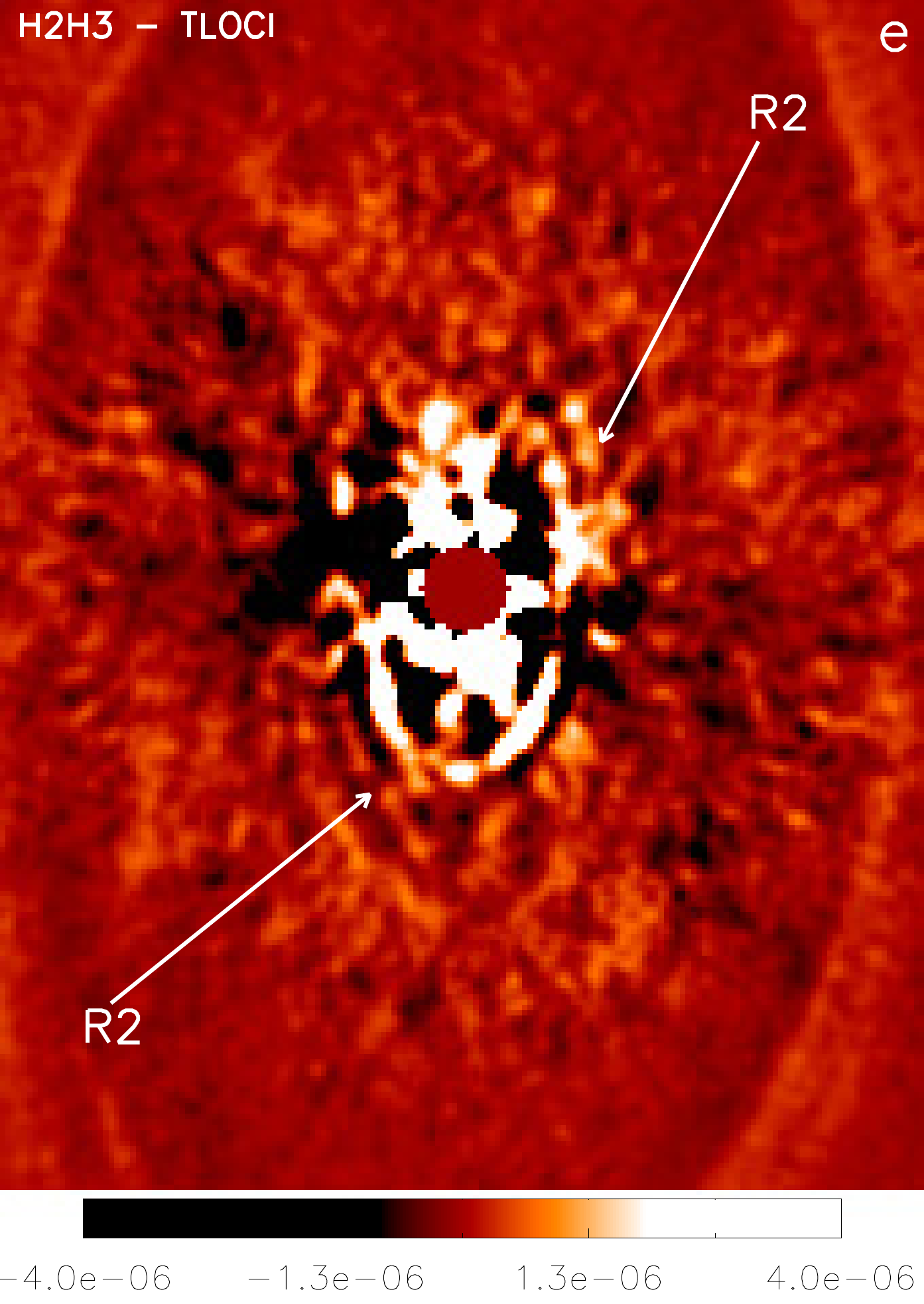}
\includegraphics[width=5.5cm]{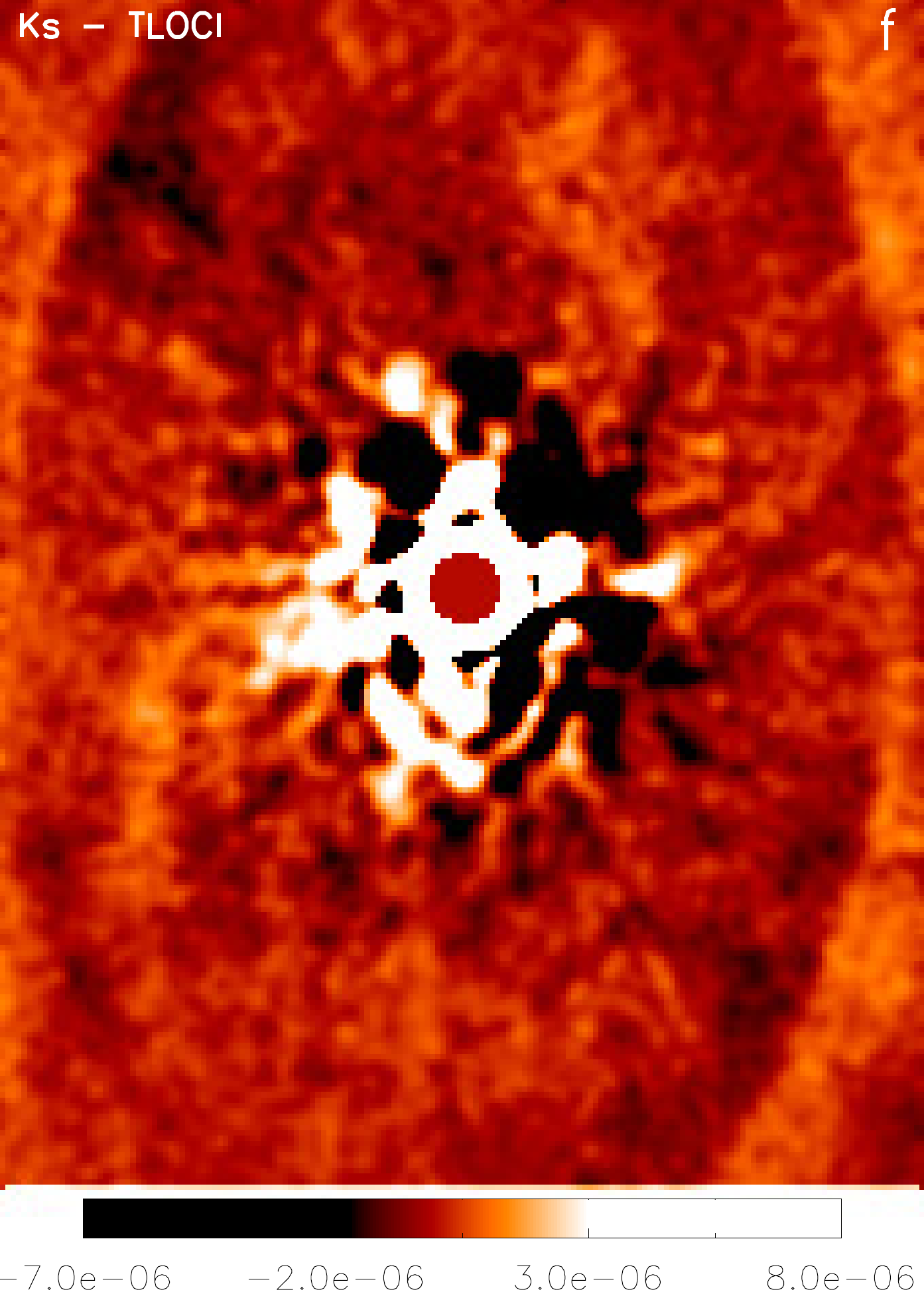}\\
\includegraphics[width=5.5cm]{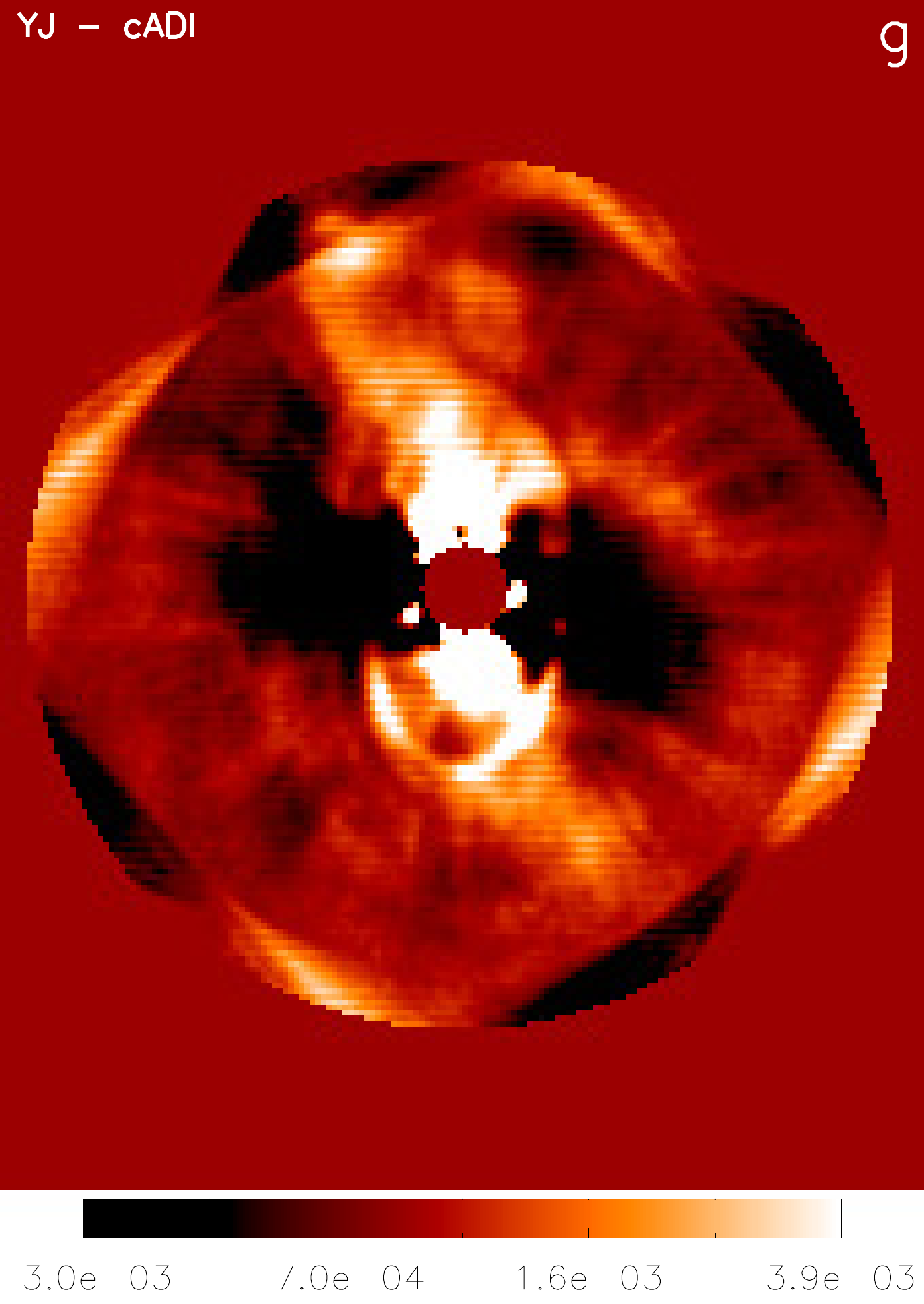}
\includegraphics[width=5.5cm]{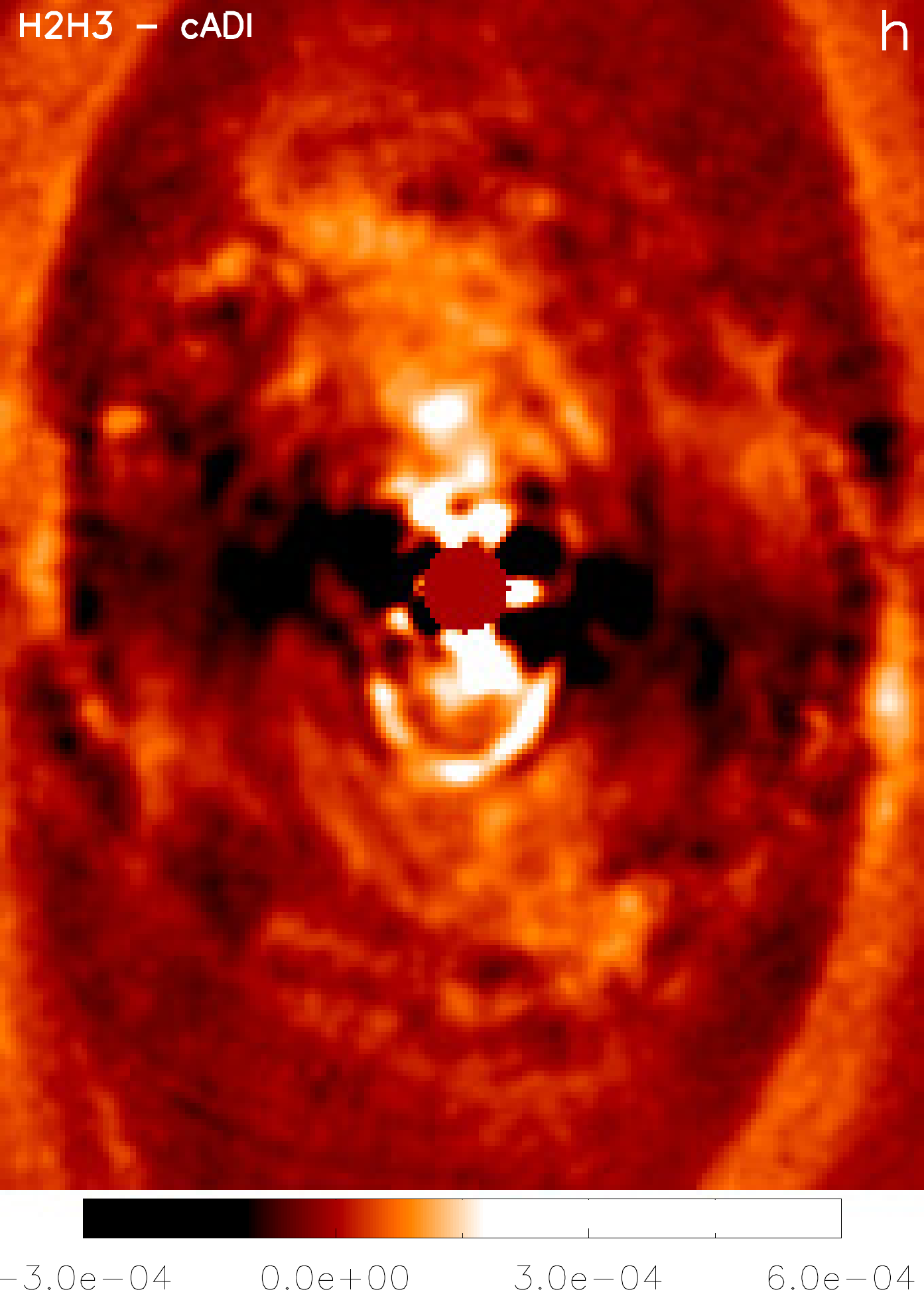}
\includegraphics[width=5.5cm]{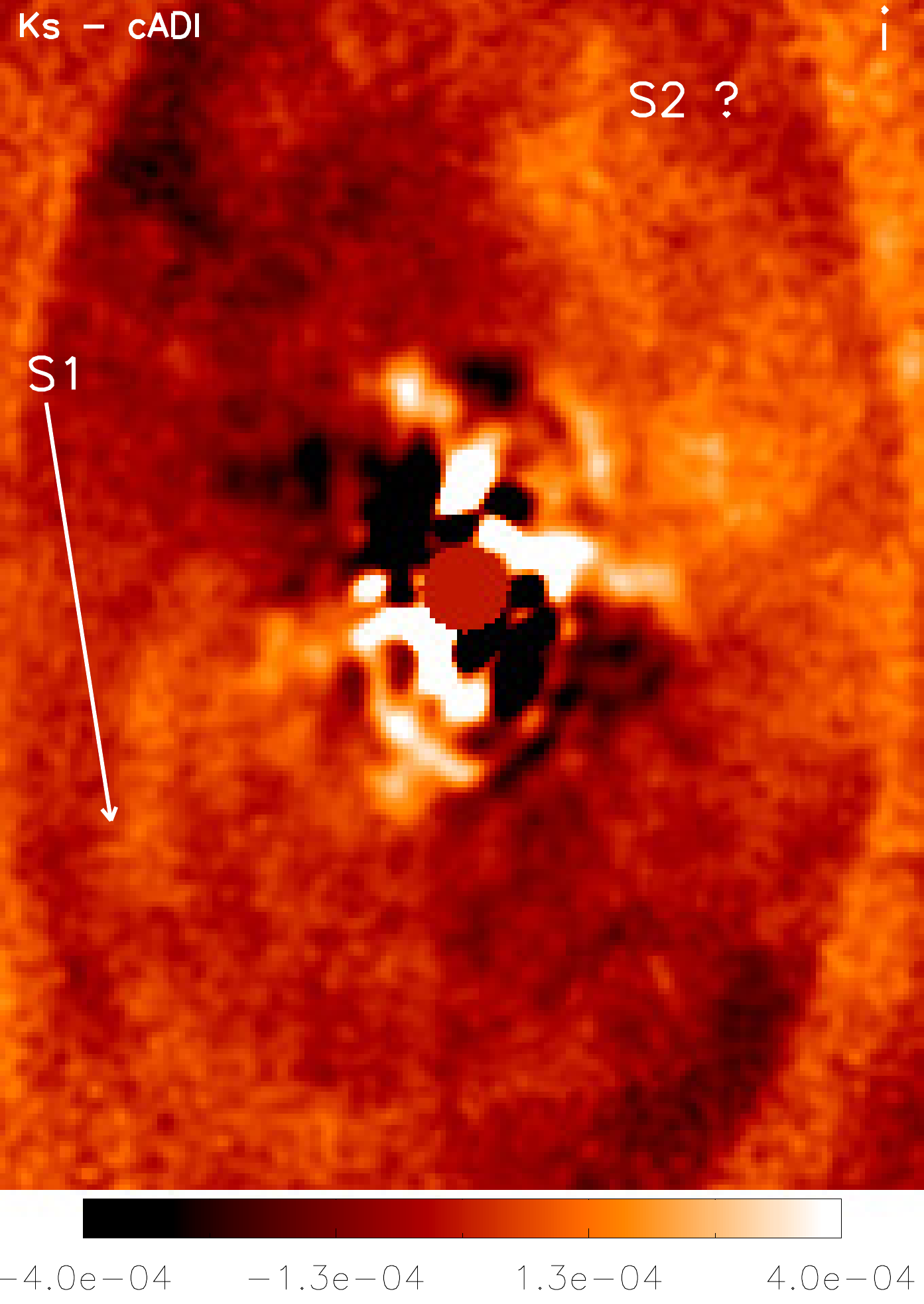}\\
\caption{Images of the central part of the system in YJ (left column), H2H3 (middle column), and Ks (right column) bands with three differents post-processing : KLIP (top row), TLOCI (middle row), and cADI (bottom row). The cADI images are multipled by $r$ to improve the visibility of structures. The annotations indicate structures R1, R2, R3, S1-S2, and the clump for a better identification..}
\label{imgannot}
\end{figure*}

\begin{figure*}[h!]
\centering
\includegraphics[width=18cm]{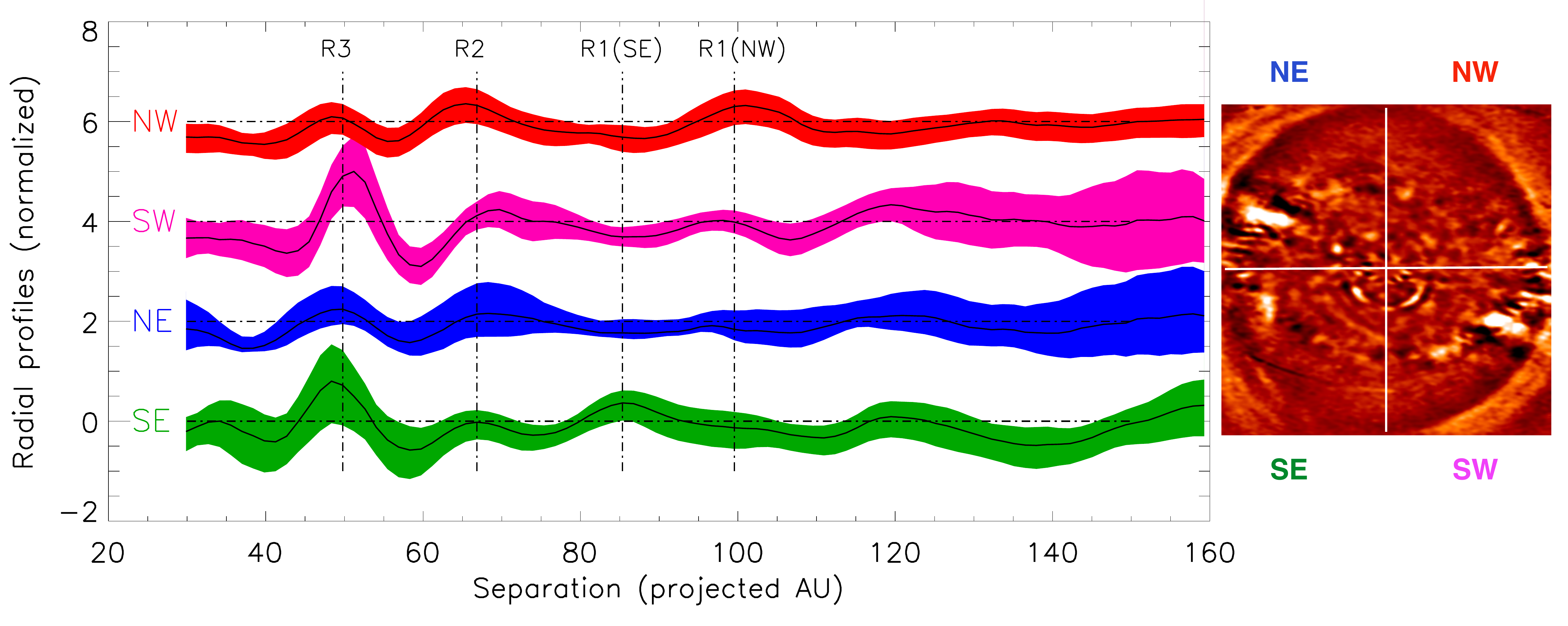}
\caption{Radial profiles of the KLIP IRDIS-H2H3 deprojected image as measured in four quadrants shown in the right panel (red : north-west, magenta : south-west, blue : north-east and green : south-east). Black lines stand for averaged profile and the colour shaded areas indicate the azimuthal dispersion. All profiles are normalized and vertically shifted for the sake of clarity}
\label{coupe}
\end{figure*}

\begin{figure*}[h!]
\centering
\includegraphics[width=9cm]{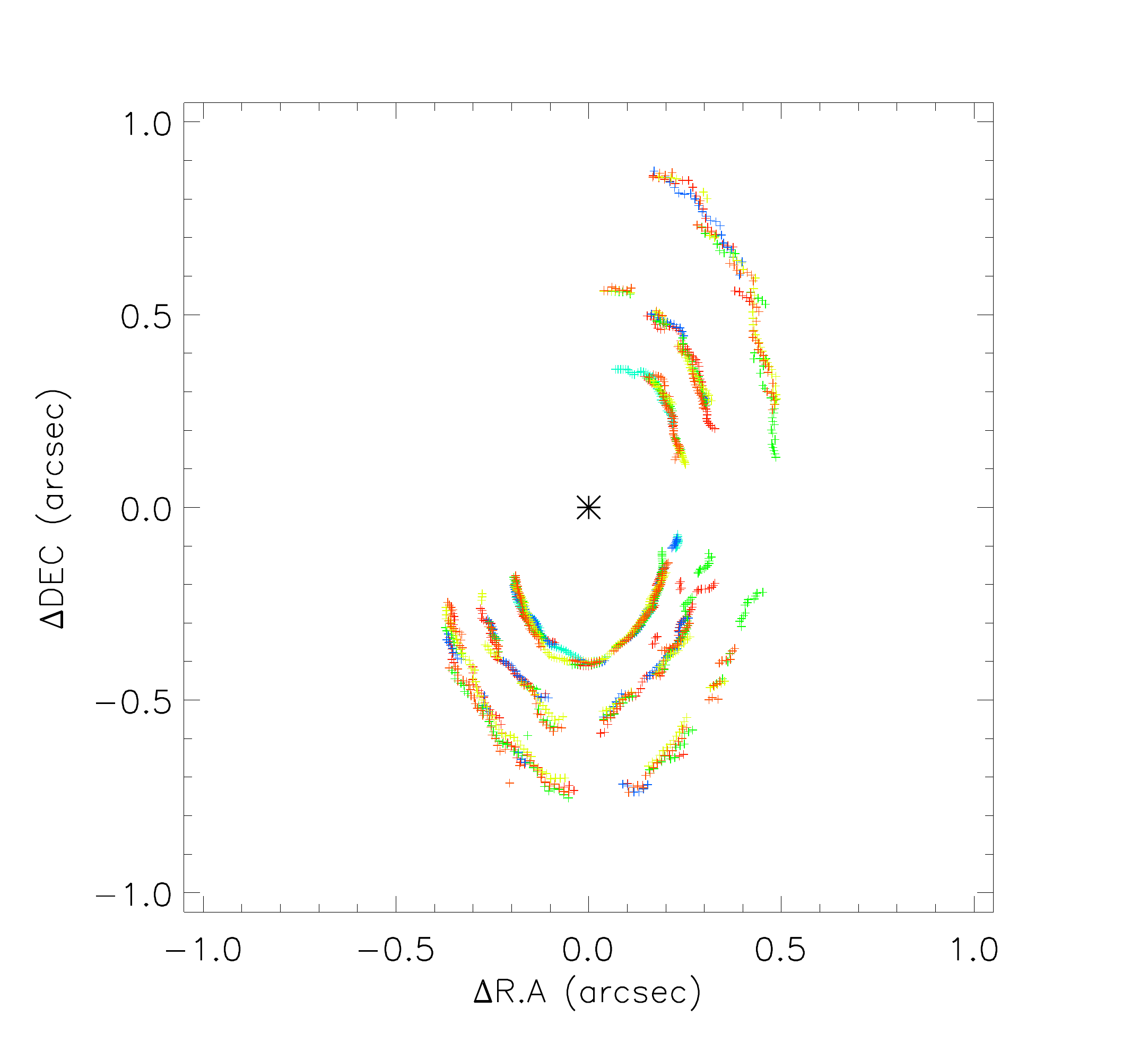}
\caption{Registration of the ringlets in the IRDIS H2H3 and Ks band images for several algorithms( KLIP, TLOCI and cADI). Ks - KLIP (red), Ks - cADI (cyan), Ks - TLOCI (blue), H2H3 - KLIP (green), H2H3 - cADI (yellow), and H2H3 TLOCI (orange). }
\label{imgmesure}
\end{figure*}

\section{Models compared to data}

\begin{figure*}[h!]
\centering
\includegraphics[width=18cm]{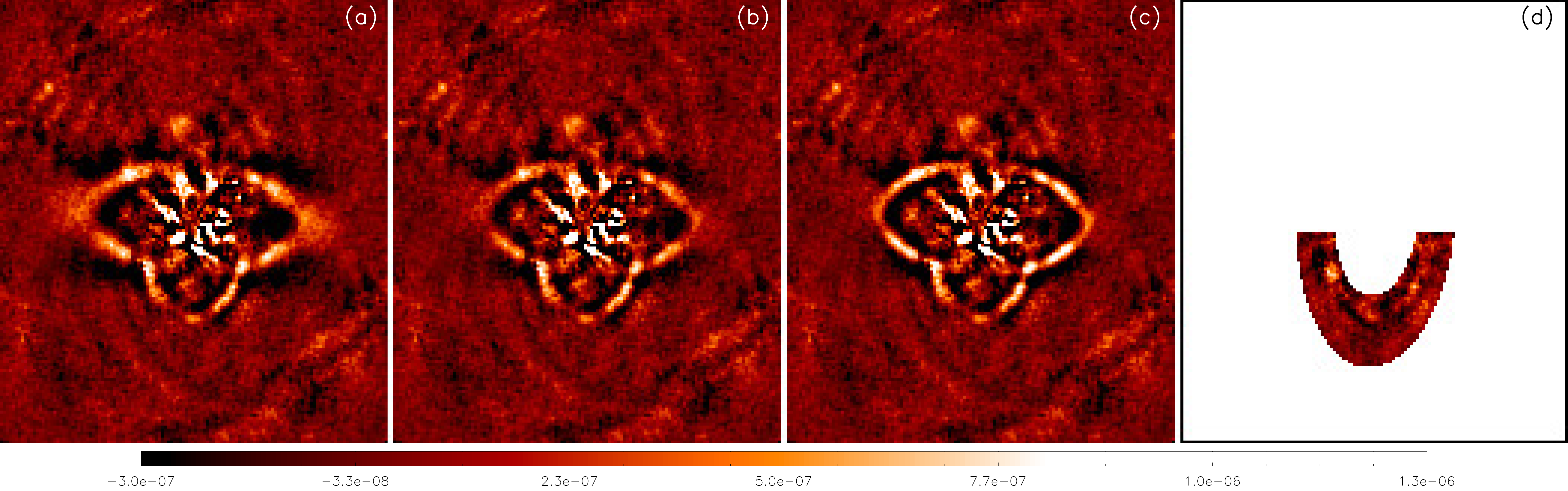}
\caption{Comparison between the H2H3 image (vertical component) and disk models (horizontal component) with different values of $\alpha_{in}$ and $\alpha_{out}$. a : $\alpha_{in} = 5$ and $\alpha_{out} = -5$ ; b : $\alpha_{in} = 10$ and $\alpha_{out} = -10$ ; c : $\alpha_{in} = 20$ and $\alpha_{out} = -20$. The inclination is set to 58\deg and the semi-major axis is set to 0.41''. Panel d shows the residuals after subtraction of the best model ($\alpha_{in} = 20$ and $\alpha_{out} = -20$).}
\label{model}
\end{figure*}

\section{Contrast limits for point source}

\begin{figure*}[h!]
\centering
\includegraphics[width=10cm]{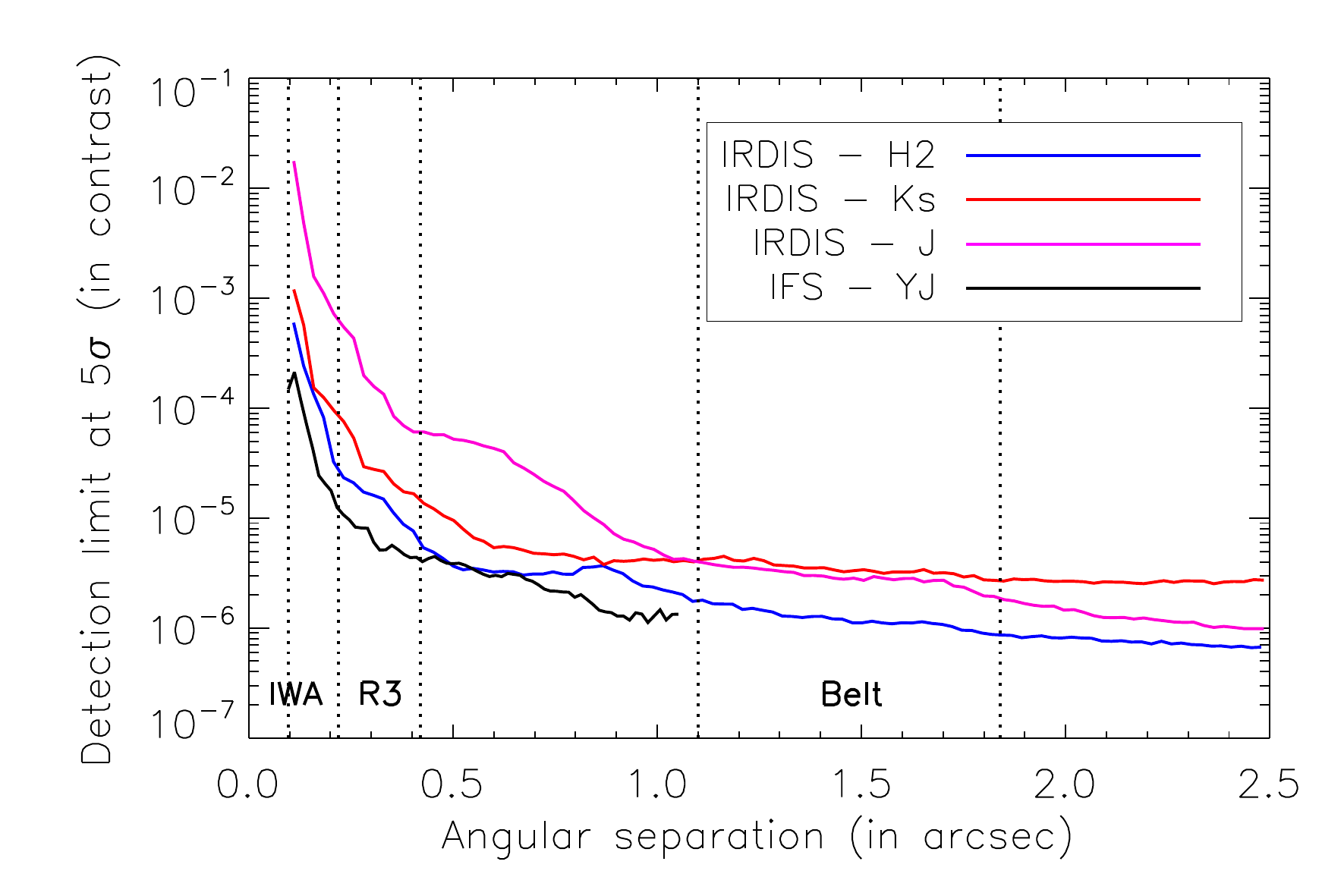}\\
\caption{Detection limit in contrast for YJ, J, H2, and Ks bands with the TLOCI images. The contrast is obtained by an azimuthal standard deviation for each angular separation, corrected by the throughput (\texttt{SpeCal}, R. Galicher, private communication).}
\label{cstcurve1}
\end{figure*}

\end{document}